\documentclass[journal=gmj]{CUP-JNL-DTM}%
\usepackage[utf8]{inputenc}

\usepackage{graphicx}
\graphicspath{{figures/}}
\usepackage{multicol,multirow}
\usepackage{amsmath,amssymb,amsfonts}
\usepackage{mathrsfs}
\usepackage{amsthm}
\usepackage{rotating}
\usepackage{appendix}
\usepackage{ifpdf}
\usepackage[T1]{fontenc}
\usepackage{algpseudocode}
\usepackage{algorithm}
\usepackage{newtxtext}
\usepackage{newtxmath}
\usepackage{textcomp}
\usepackage{xcolor}
\usepackage{lipsum}
\usepackage[colorlinks,allcolors=blue]{hyperref}
\usepackage{siunitx}
\theoremstyle{definition}

\numberwithin{equation}{section}

\jname{SAMBoost}
\articletype{RESEARCH PAPER}
\makeatletter
\def\ps@titlepage{\let\@oddhead\@empty\let\@evenhead\@empty\let\@evenfoot\@oddfoot}
\makeatother

\usepackage[utf8]{inputenc}
\usepackage{float}
\usepackage{hyperref}
\usepackage{graphicx}
\usepackage{makecell}
\usepackage{enumitem}

\usepackage{geometry}
\usepackage{amssymb}
\usepackage{caption}
\usepackage{subcaption}
\usepackage{cmll}
\usepackage{bbm}
\usepackage{array, booktabs, boldline} 
\usepackage{amsmath}
\usepackage{changepage}

\usepackage[textwidth=1.8cm, textsize=tiny]{todonotes}

\begin{document}

\begin{Frontmatter}

\title[SAMBoost]{SAMBoost: Spatially Aware Mesh Boosting}

\author[1,2]{Mulah Moriah}
\author[1,2]{Franck Vermet}
\author[1,2]{Pierre Ailliot}
\author[1,2]{Juliette Legrand}

\authormark{Mulah Moriah \textit{et al}.}

\address[1]{\orgdiv{Euro-Institut d'Actuariat Jean Dieudonné (EURIA)}, \orgname{UBO}, \orgaddress{\city{Brest}, \country{France}}}

\address[2]{\orgdiv{Univ Brest, CNRS UMR 6205}, \orgname{Laboratoire de Mathématiques de Bretagne Atlantique, Brest}, \orgaddress{\country{France}}}

\address{\textbf{Corresponding author: }Mulah Moriah; Email: mulah.moriah@etudiant.univ-brest.fr}

\keywords{gradient boosting, spatial random effects, SPDE, Gaussian Markov random fields, flood risk, insurance}

\abstract{We propose SAMBoost (Spatially Aware Mesh Boosting), a gradient boosting method that combines tree-based covariate modelling with a stochastic partial differential equation (SPDE) latent spatial field through a single penalised objective. At each boosting iteration, the predictor is decomposed into separate covariate and spatial components. The method is motivated by applications in which the spatial effect is not a nuisance to be removed but a quantity of interest that must be recovered alongside the covariates. In insurance, for example, practitioners need both flexible nonlinear covariate effects and an interpretable geographic risk surface that can be reported on its own. Existing approaches often involve trade-offs, offering flexible covariate modelling but no explicit spatial field, rigorous spatial inference with a fixed additive covariate structure, or a combination of both at computational costs that limit practicality for large datasets. SAMBoost addresses these goals within the boosting loop, and the underlying mesh produces a sparse precision structure whose cost is largely independent of sample size. On simulations and on an open benchmark, SAMBoost matches or improves on the predictive accuracy of joint Gaussian process and tree boosting approaches. It is also faster and recovers the spatial field and covariate effects more accurately. When covariate effects include interactions, it outperforms additive Bayesian spatial models. Applied to a flood insurance portfolio with 4.9 million records, SAMBoost trains on the full training set without subsampling and runs about sixteen times faster than a Bayesian reference, with a consistent spatial risk surface. A public Python implementation is available.}

\end{Frontmatter}

\section{Introduction}
\subsection{Spatial dependence and the attribution problem}\label{sec:intro-spatial}
 
Tobler's first law of geography states that everything is related to everything else, but near things are more related than distant things~\citep{Tobler1970}. In practice, this means that both the outcome of interest and the covariates used to predict it tend to vary over space. When a model does not account for this spatial dependence, covariates exhibiting spatial structure are forced to absorb a geographic signal that does not belong to them, distorting both the predictive surface and the interpretation of individual covariate effects~\citep{HodgesReich2010,Sigrist2022,SahaBasuDatta2023,Moriah2026}. This has motivated a substantial body of work on the explicit integration of spatial structure into regression and prediction models, which has proceeded in two broad directions.
 
The first direction, rooted in spatial statistics, models the geographic component as a latent random effect estimated jointly with the covariates. Conditional autoregressive priors~\citep{GschloslCzado2007}, penalised spline structures within Bayesian generalised additive models~\citep{DenuitLang2004}, and the stochastic partial differential equation (SPDE) representation of Matérn fields~\citep{LindgrenRueLindstrom2011} all produce a spatial field that can be reported separately from the covariate effects, with full uncertainty quantification. However, the covariate part of the model must be pre-specified as a linear predictor or a low-dimensional sum of smooth functions, which struggles to capture the high-dimensional nonlinear interactions present in observational datasets. 

The second direction integrates spatial information into flexible machine learning estimators. Gradient boosted trees with coordinate features~\citep{HenckaertsCoteetal2021}, geographically weighted approaches and their nonlinear extensions~\citep{Fotheringhametal2002,Grekousis2025}, and spatial loss regularisation~\citep{Geertsetal2024} all offer predictive flexibility over large feature spaces, but do not maintain an explicit latent spatial field during training. 

The entanglement of spatial and covariate signals within these models is closely related to the well-studied problem of spatial confounding~\citep{Paciorek2010,HodgesReich2010}, which has received extensive treatment in the linear and generalised additive model literature~\citep{DupontMarquesKneib2025}. However, that line of work is concerned with protecting covariate effect estimates from spatial bias rather than with constructing and extracting the spatial field as an independent modelling object, which is the focus of the present paper.
 
The two directions of research each address one of the two demands that spatially structured data create. The first is a predictive demand for flexible nonlinear covariate modelling. The second is an attributive demand for a geographic effect that, while estimated jointly with the covariates, can be cleanly separated and independently reported. We propose SAMBoost (Spatially Aware Mesh Boosting) as the bridge between these two traditions: an algorithm that satisfies both demands simultaneously within a unified estimation loop while remaining tractable for large datasets and fine-grained spatial resolutions encountered in applied practice. 

This dual requirement arises in a concrete way in insurance risk modelling, especially in property insurance for climate-related hazards, where spatial clustering of claims reflects shared environmental exposure rather than only the observed rating variables. Modern pricing approaches combine diverse data sources (building attributes, weather records, institutional hazard maps) whose spatial patterns overlap and interact. The insurer must therefore be able to model this in a way that distinguishes the variation explained by observable covariates from that arising from an unobserved spatial component. The demand for high spatial resolution is particularly pronounced for flood risk, where sharp risk gradients appear below the municipal level, and coarse territorial classifications are known to distort risk analysis~\citep{Moriah2026}. These applied 
considerations have guided the design of SAMBoost and motivate the insurance application discussed in Section~\ref{sec:application}.

The broader spatial machine learning literature has engaged with this dual challenge from several architectural angles, and we review it in the next section to expose the precise methodological gap that SAMBoost addresses.

\subsection{Spatial machine learning methods: review and positioning}\label{sec:intro-review}

Beyond the broad characterisation of the previous section, a substantial spatial machine learning literature has developed methods for learning from spatially correlated data. These methods can be organised into five families according to the architectural role they assign to the spatial component.\vspace{3mm}

\textbf{Spatial feature augmentation.} \noindent The simplest strategy appends coordinate-derived features to the covariate matrix and fits a standard learner on the augmented input~\citep{Hengletal2018,HenckaertsCoteetal2021}. When the learner is a tree-based model, the feature space is partitioned with axis-aligned splits, producing spatial surfaces that are discontinuous and exhibit rectangular artefacts. More fundamentally, geographic location enters the model through the same mechanism as any other covariate, so the spatial and covariate contributions are entangled inside every split and cannot be separated after fitting.

\textbf{Post-hoc spatial correction.} A second approach fits a machine learning model under the independence assumption and then applies a geostatistical smoother to its residuals~\citep{Hengletal2018,xiao2023mapping}. This produces a spatial surface, but one that is entirely conditional on whatever signal the machine learning model has already absorbed. If the covariate model has captured part of the geographic variation, the residual surface reflects what remains rather than the true spatial component, and the degree of contamination depends on the model's effective capacity.

\textbf{Local regression methods.} Geographically weighted regression~\citep{Fotheringhametal2002} and its extensions to random forests~\citep{Georganos2021} and gradient boosting~\citep{Wang2026,Grekousis2025} allow covariate effects to vary across space by fitting local models within spatial neighbourhoods. These methods capture spatial heterogeneity in the covariate relationships, but conflate the spatial structure with the covariate effects within each local model: the geographic variation is expressed through spatially varying coefficients rather than through an independent spatial field.

\textbf{Spatial loss regularisation.} Geerts et al. \cite{Geertsetal2024} penalise the entire predicted surface for spatial autocorrelation within the boosting loss function. This forces both the covariate and geographic components to satisfy the same smoothness constraint, which is a severe restriction when covariate effects exhibit sharp nonlinear transitions that should not be smoothed.

These four families share a common structural limitation: none produces an explicit latent spatial field that is maintained and updated as a modelling object during training.

\vspace{2mm}
\textbf{Joint machine learning and spatial field estimation.} The fifth family explicitly models the spatial correlation structure as a latent random effect and integrates it with a tree-based estimator during training.
Saha et al.~\citep{SahaBasuDatta2023} proposed RF-GLS, which replaces the ordinary least squares split criterion in random forests with a generalised least squares objective calibrated to the Gaussian process (GP) covariance structure.
Huang et al.~\citep{HuangMcMahanetal2025} and MacBride et al.~\citep{MacBrideetal2025} both combine iterative tree-based estimation with a conditional autoregressive spatial model, the former applied to disease surveillance and the latter to small-area prediction.
Hu et al.~\citep{Huetal2026} take a sequential approach in which XGBoost forecasts serve as covariates within a latent Gaussian model fitted by integrated nested Laplace approximation (INLA) for wildfire prediction; the spatial field in the second stage is well-defined but is not updated jointly with the tree ensemble.
A particularly relevant contribution is GPBoost~\citep{Sigrist2022,Sigrist2023}, which jointly trains a tree ensemble and a Gaussian process random effect within a single likelihood framework.
At each boosting round, GPBoost passes a gradient decorrelated by generalised least squares to the tree, removing spatial redundancy from the pseudo-residuals before tree fitting.
The spatial field can be extracted after training via posterior kriging of the final residuals.

Although these methods represent a substantial advance toward joint estimation, two limitations remain.
First, the spatial field is not structurally separated from the covariate component during training: it is either recovered from residuals after fitting or absorbed into a joint likelihood where the tree and the spatial effect compete for the same signal without an explicit separation mechanism.
Second, as empirically demonstrated by Wong et al.~\citep{Wongetal2023}, methods that operate directly on observation-level covariance structures do not scale well as the number of spatial locations increases, becoming computationally infeasible for datasets with more than $10^4$ unique locations.

SAMBoost positions itself within this fifth family, sharing its commitment to explicit latent spatial modelling and joint estimation, while addressing both limitations. The next section details these contributions and outlines the remainder of the paper.

\subsection{Contributions and paper organisation}\label{sec:contributions}

The central contribution of this work is the design of a boosting procedure in which the spatial field is not a post-hoc artefact of the fitted model but an object that is built up progressively across boosting rounds, structurally separated from the covariate component at every iteration. At each round, the algorithm performs two operations. The first fits a regression tree to the current pseudo-residuals using only the covariate matrix, with no geographic information passed to the tree. The second updates the spatial field by solving a sparse penalised weighted least squares system on a triangulated mesh derived from the SPDE representation of a Matérn Gaussian field~\citep{LindgrenRueLindstrom2011}, using only the current residual and the mesh geometry. Because the two operations receive structurally separated inputs at every round, the final predictor admits an exact additive decomposition into a covariate component that depends only on the covariates and a spatial component that depends only on geographic location. When the spatial field itself is a quantity of interest, it deserves the same modelling attention as the covariate component. SAMBoost reflects this by building the spatial field progressively across boosting rounds, giving it the same iterative refinement that gradient boosting affords to the covariate model. This contrasts with GPBoost~\citep{Sigrist2022}, where the spatial component serves primarily as a debiasing device for the tree ensemble and is recovered after kriging the final residuals.

The second contribution concerns computational tractability. Representing the spatial field on a triangulated SPDE mesh rather than directly at observation locations yields a sparse precision matrix whose Cholesky factorisation scales as $\mathcal{O}(M^{1.5})$ for a mesh of $M$ nodes. The spatial update thus stays cheap even when $M$ is close to the number of observations $N$, as in our simulations, and $M$ becomes much smaller than $N$ on large datasets. The mesh construction algorithm seeds nodes from Poisson-disc-filtered observation locations and boundary hull nodes, ensuring that the mesh resolution adapts to the spatial density of the data. This makes SAMBoost tractable at fine-grained spatial resolutions where observation-level methods become computationally infeasible~\citep{Wongetal2023}, and avoids the dependence on administrative zoning that limits methods relying on areal spatial units.

We empirically compare SAMBoost with two methods that, like SAMBoost, deliver an explicit spatial field while occupying distinct positions in this methodological space. GPBoost~\citep{Sigrist2022} is the most directly comparable approach, as it likewise performs joint estimation of tree-based effects and spatial structure within a single likelihood framework. INLA with an SPDE-based spatial field~\citep{LindgrenRueLindstrom2011,Moriah2026} serves as the spatial statistics reference, offering rigorous inference but under a pre-specified additive covariate model. Feature-augmentation approaches, which provide no explicit spatial field, are discussed separately in Section~\ref{sec:insurance}.
To the best of our knowledge, SAMBoost is the only approach that simultaneously offers a flexible nonlinear covariate model, retains an explicit latent spatial field throughout training, yields an extractable spatial component that is structurally disentangled from the covariates, and remains computationally feasible when the number of spatial locations exceeds $10^4$.

The remainder of the paper is organised as follows. Section~\ref{sec:model} introduces the model and the two-stage boosting algorithm. Section~\ref{sec:simulations} presents simulated experiments that benchmark SAMBoost against the two baseline models. Section~\ref{sec:application} applies the method to an open house price benchmark and to a national flood insurance portfolio. Section~\ref{sec:discussion} concludes with a discussion of limitations and future directions.

\section{Model and Algorithm}\label{sec:model}

\subsection{Model specification}\label{sec:model-spec}

We consider the problem of learning a predictive model for an outcome $y_i$ observed at $N$ locations, where each observation is associated with a covariate vector $x_i \in \mathbb{R}^p$ and a spatial position $s_i \in \mathbb{R}^2$. The goal is to jointly estimate a flexible nonlinear covariate effect and an explicit spatial field, with both components separately identifiable and extractable after training. To this end, we model the response through an additive predictor:
\begin{equation}\label{eq:predictor}
F_i \;=\; F_0 \;+\; F_{\mathrm{cov}}(x_i) \;+\; F_{\mathrm{spa}}(s_i)\,.
\end{equation}
Here, $F_0$ denotes a fixed intercept, $F_{\mathrm{cov}}\colon \mathbb{R}^p \to \mathbb{R}$ represents the covariate effect, and $F_{\mathrm{spa}}\colon \mathbb{R}^2 \to \mathbb{R}$ represents the spatial effect. The link between $F_i$ and the response $y_i$ is defined via a loss function $\ell(y_i, F_i)$, which is assumed to be convex and twice differentiable with respect to its second argument. This assumption holds for squared error and logistic losses \citep{Friedman2001}, and for Poisson losses with a log link, which are common in insurance \citep{HenckaertsCoteetal2021}.

The covariate component adopts the standard gradient boosting framework~\citep{Friedman2001}. After $T$ boosting iterations it is given by
\begin{equation}\label{eq:fcov}
F_{\mathrm{cov}}^{(T)}(x) \;=\; \sum_{t=1}^{T} \eta\, f_t(x)\,,
\end{equation}
where each base learner $f_t$ is a shallow regression tree and $\eta \in (0,1)$ denotes the learning rate. The trees take only the covariate matrix $\mathbf{X}$ as input, and no spatial information is supplied to the tree learner at any stage. Because the covariate update requires only one regression tree per iteration, any compatible tree implementation (exact or histogram-based) can be used as a plug-in backend without modifying the overall approach.

The spatial component is represented on a triangulated mesh with $M$ nodes covering the study region. Concretely, 
\begin{equation}\label{eq:fspa-projection}
F_{\mathrm{spa}}(s_i) = [\mathbf{A}\mathbf{z}]_i\,,
\end{equation}
where $\mathbf{z} \in \mathbb{R}^M$ stores the spatial field values at the mesh nodes and $\mathbf{A} \in \mathbb{R}^{N \times M}$ is a projection matrix that transfers these values to the observation points using barycentric interpolation. For each observation $s_i$, the three vertices of its containing triangle contribute weights that sum to one, producing at most three nonzero entries per row of $\mathbf{A}$. The matrix is assembled once before training and reused at every iteration. The mesh itself is constructed from the observation locations using Delaunay triangulation with Poisson-disc thinning and boundary extension, following the SPDE mesh framework of \citep{LindgrenRueLindstrom2011}. By design, interior mesh nodes are placed at data locations, so the resolution adapts to the local density of observations. The number of nodes $M$ is therefore close to $N$ on small datasets, such as our simulations, but much smaller on large ones.

Like the covariate component, the spatial field is built up incrementally over the boosting iterations as
\begin{equation}\label{eq:zspatial}
\mathbf{z}^{(T)} \;=\; \sum_{t=1}^{T} \eta\,\Delta\mathbf{z}^{(t)}\,,
\end{equation}
where $\Delta\mathbf{z}^{(t)}$ is the spatial increment at round $t$, scaled by the same learning rate $\eta$. Because $F_{\mathrm{cov}}$ depends only on $x_i$ and $F_{\mathrm{spa}}$ depends only on $s_i$, the additive predictor admits an exact decomposition into a covariate contribution and a spatial contribution at every observation, by construction.

\subsection{Penalised objective and Newton derivation}\label{sec:objective}

The algorithm is built around a penalised loss of the form
\begin{equation}\label{eq:objective}
\mathcal{L}(F_{\mathrm{cov}}, \mathbf{z}) \;=\; \sum_{i=1}^{N} \ell(y_i, F_i) \;+\; \frac{\lambda_{\mathrm{tree}}}{2}\sum_{t=1}^{T}\sum_{k=1}^{J_t} c_{tk}^2 \;+\; \frac{\lambda_{\mathrm{spa}}}{2}\,\mathbf{z}^\top \mathbf{Q}\,\mathbf{z}\,,
\end{equation}
where $\ell(y_i, F_i)$ is a convex, twice-differentiable loss function (e.g.\ logistic, squared error, or Poisson deviance). The objective combines three terms. The first is the data-fitting loss summed over all observations. 

The second term imposes $L_2$ regularisation on the leaf values $c_{tk}$ of tree $t$ ($k = 1,\ldots,J_t$ leaves), with $\lambda_{\mathrm{tree}} > 0$ controlling the shrinkage, consistent with XGBoost~\citep{ChenGuestrin2016}.

The third term penalises roughness in the spatial field $\mathbf{z}$. It is equivalent to placing a zero-mean Gaussian Markov random field (GMRF) prior on $\mathbf{z}$ with precision matrix $\lambda_{\mathrm{spa}}\,\mathbf{Q}$~\citep{rue2005gaussian}, so that $\mathbf{Q}$ encodes how strongly neighbouring mesh nodes are tied together. We take this prior to be the mesh approximation of a Gaussian field $u(s)$ with Mat\'ern covariance, the standard covariance family in spatial statistics. For smoothness $\nu = 1$ in two dimensions, such a field is the solution of the SPDE~\citep{LindgrenRueLindstrom2011}
\begin{equation}\label{eq:spde}
(\kappa^2 - \Delta)\,\tau\, u(s) = \mathcal{W}(s)\,,
\end{equation}
where $\Delta$ is the Laplacian, $\mathcal{W}$ is spatial white noise, $\kappa = \sqrt{8}/\rho$ sets the practical range $\rho$ (the distance at which the correlation falls to about 0.1), and $\tau$ scales the marginal variance $\sigma^2 = 1/(4\pi\,\kappa^2\,\tau^2)$. Discretising this equation on the triangulated mesh, with $\mathbf{z}$ the values of $u$ at the mesh nodes, yields
\begin{equation}\label{eq:precision}
\mathbf{Q} = \tau^2\bigl(\kappa^4\,\mathbf{C} + 2\kappa^2\,\mathbf{G} + \mathbf{G}\,\mathbf{C}^{-1}\,\mathbf{G}\bigr)\,,
\end{equation}
where $\mathbf{C}$ is the diagonal mass matrix and $\mathbf{G}$ is the stiffness matrix of the mesh~\citep{LindgrenRueLindstrom2011}. Both are sparse, so $\mathbf{Q}$ has nonzero entries only between nodes at most two edges apart and its Cholesky factorisation costs $\mathcal{O}(M^{1.5})$, instead of $\mathcal{O}(N^3)$ for the dense covariance of a Gaussian process.

\noindent The parameters $\kappa$ and $\tau$ are held fixed throughout the boosting procedure. Since $\mathbf{Q}$ depends on $\tau$, only the product $\lambda_{\mathrm{spa}}\,\tau^2$ affects the penalty. We therefore fix $\tau = 1/(\sqrt{4\pi}\,\kappa)$, which normalises the marginal variance to $\sigma^2 = 1$, and let $\lambda_{\mathrm{spa}}$ absorb all penalty scaling. This leaves $\kappa$ (equivalently the practical range $\rho$) and $\lambda_{\mathrm{spa}}$ as the two SPDE-related tuning parameters.

The additive form of the predictor, $F_i = F_0 + F_{\mathrm{cov}}(x_i) + F_{\mathrm{spa}}(s_i)$, suggests an alternating optimisation strategy. In each boosting iteration, the algorithm first updates the covariate term while keeping the spatial field fixed, and then updates the spatial field while holding the covariate term constant. Both updates are obtained from a second-order Taylor expansion of the objective~\eqref{eq:objective}, restricted to the relevant block of parameters. We detail the derivation of each step in the following. \vspace{3mm}

\textbf{Stage~1, the Newton leaf update.} At round $t$, the structure of the tree $f_t$ is grown by a least-squares fit to the negative gradient, as in the TreeBoost scheme of \citep{Friedman2001}, and partitions the observations into $J$ disjoint leaves $\mathcal{R}_1, \ldots, \mathcal{R}_J$. Given this partition, the leaf values are set by a regularised Newton step. Within leaf $\mathcal{R}_k$, the tree assigns a constant value $c_k$, altering the predictor by $\delta_i = c_k$ for all $i \in \mathcal{R}_k$. Let $g_i = \partial\ell/\partial F\big|_{F_i}$ and $h_i = \partial^2\ell/\partial F^2\big|_{F_i}$ denote the gradient and Hessian of the loss at the current predictor. A second-order Taylor expansion of the loss around $F_i$ gives
\begin{equation}\label{eq:taylor_leaf}
\ell(y_i, F_i + c_k) \;\approx\; \ell(y_i, F_i) \;+\; g_i\, c_k \;+\; \frac{h_i}{2}\, c_k^2\,.
\end{equation}
Summing over all observations in $\mathcal{R}_k$ and letting $G_k = \sum_{i \in \mathcal{R}_k} g_i$ and $H_k = \sum_{i \in \mathcal{R}_k} h_i$, and further incorporating the $L_2$ regularisation on $c_k$, the resulting change in the penalised objective, restricted to leaf $k$, satisfies
\begin{equation}\label{eq:quad_leaf}
\sum_{i \in \mathcal{R}_k}\bigl[\ell(y_i, F_i + c_k) - \ell(y_i, F_i)\bigr] + \frac{\lambda_{\mathrm{tree}}}{2}\,c_k^2 \;\approx\; G_k\, c_k \;+\; \frac{H_k + \lambda_{\mathrm{tree}}}{2}\, c_k^2\,.
\end{equation}
Differentiating with respect to $c_k$ and setting the derivative to zero gives the Newton update for the leaf value
\begin{equation}\label{eq:newton_leaf}
\hat{c}_k \;=\; -\,\frac{G_k}{H_k + \lambda_{\mathrm{tree}}}\,.
\end{equation}
The leaf value is therefore the Hessian-weighted average of the per-observation Newton steps $-g_i/h_i$ within the leaf, shrunk toward zero by the regularisation parameter $\lambda_{\mathrm{tree}}$. The resulting tree prediction $f_t(x_i) = \hat{c}_k$ for all $i \in \mathcal{R}_k$ is then multiplied by the learning rate $\eta$ when updating the model: $F_{\mathrm{cov}} \leftarrow F_{\mathrm{cov}} + \eta\, f_t(x)$ (see \cite{ChenGuestrin2016}, the original XGBoost paper, for further details). \vspace{3mm}

\textbf{Stage~2, the Newton spatial update.} Once Stage~1 has updated the predictor, the gradient $g_i$ and Hessian $h_i$ are recalculated at the new values of $F_i$. The spatial increment $\Delta\mathbf{z}$ modifies the predictor via $\delta_i = [\mathbf{A}\Delta\mathbf{z}]_i$ (recall from~\eqref{eq:fspa-projection} that $F_{\mathrm{spa}} = \mathbf{A}\mathbf{z}$). Using a second-order Taylor expansion of the data loss and summing over all $N$ observations yields
\begin{equation}\label{eq:taylor_spatial}
\sum_{i=1}^{N}\bigl[\ell(y_i, F_i + [\mathbf{A}\Delta\mathbf{z}]_i) - \ell(y_i, F_i)\bigr] \;\approx\; \mathbf{g}^\top\mathbf{A}\Delta\mathbf{z} \;+\; \frac{1}{2}(\mathbf{A}\Delta\mathbf{z})^\top\mathbf{H}(\mathbf{A}\Delta\mathbf{z})\,,
\end{equation}
where $\mathbf{g} = (g_1, \ldots, g_N)^\top$ and $\mathbf{H} = \operatorname{diag}(h_1, \ldots, h_N)$. The change in the GMRF penalty when updating from $\mathbf{z}$ to $\mathbf{z} + \Delta\mathbf{z}$ is
\begin{equation}\label{eq:penalty_change}
\frac{\lambda_{\mathrm{spa}}}{2}(\mathbf{z} + \Delta\mathbf{z})^\top\mathbf{Q}(\mathbf{z} + \Delta\mathbf{z}) - \frac{\lambda_{\mathrm{spa}}}{2}\mathbf{z}^\top\mathbf{Q}\mathbf{z} \;=\; \lambda_{\mathrm{spa}}\,\mathbf{z}^\top\mathbf{Q}\,\Delta\mathbf{z} \;+\; \frac{\lambda_{\mathrm{spa}}}{2}\Delta\mathbf{z}^\top\mathbf{Q}\,\Delta\mathbf{z}\,.
\end{equation}
Combining~\eqref{eq:taylor_spatial} and~\eqref{eq:penalty_change}, the total second-order change in the penalised objective is the following quadratic in $\Delta\mathbf{z}$:
\begin{equation}\label{eq:quad_spatial}
\begin{aligned}
\mathcal{L}(F_{\mathrm{cov}}, \mathbf{z} + \Delta\mathbf{z}) - \mathcal{L}(F_{\mathrm{cov}}, \mathbf{z}) \;\approx\;& \bigl(\mathbf{A}^\top\mathbf{g} + \lambda_{\mathrm{spa}}\,\mathbf{Q}\mathbf{z}\bigr)^\top\Delta\mathbf{z} \\
&+\; \frac{1}{2}\Delta\mathbf{z}^\top\bigl(\mathbf{A}^\top\mathbf{H}\mathbf{A} + \lambda_{\mathrm{spa}}\,\mathbf{Q}\bigr)\Delta\mathbf{z}\,.
\end{aligned}
\end{equation}
By differentiating with respect to $\Delta\mathbf{z}$ and setting the derivative to zero, we obtain the $M$-dimensional Newton system
\begin{equation}\label{eq:newton_spatial}
\bigl(\mathbf{A}^\top\mathbf{H}\mathbf{A} + \lambda_{\mathrm{spa}}\,\mathbf{Q}\bigr)\,\Delta\mathbf{z} \;=\; -\bigl(\mathbf{A}^\top\mathbf{g} + \lambda_{\mathrm{spa}}\,\mathbf{Q}\mathbf{z}\bigr)\,,
\end{equation}
where $\mathbf{z}$ on the right-hand side denotes the accumulated spatial field, given by $\mathbf{z}^{(t-1)} = \sum_{s=1}^{t-1}\eta\,\Delta\mathbf{z}^{(s)}$ over all previous iterations, as defined in~\eqref{eq:zspatial}. This system finds the spatial increment $\Delta\mathbf{z}$ that best reduces the penalised objective in one step. The left-hand side matrix combines data curvature ($\mathbf{A}^\top\mathbf{H}\mathbf{A}$) with spatial smoothness ($\lambda_{\mathrm{spa}}\,\mathbf{Q}$), while the right-hand side balances the data-loss gradient projected onto the mesh against the penalty gradient, which shrinks the accumulated field toward the prior mean. The system matrix $\mathbf{A}^\top\mathbf{H}\mathbf{A} + \lambda_{\mathrm{spa}}\mathbf{Q}$ is symmetric positive definite and preserves the sparsity pattern of $\mathbf{Q}$, allowing the system to be solved efficiently using sparse Cholesky factorisation. \vspace{3mm}

\textit{Weighted least squares equivalence.} The Newton system~\eqref{eq:newton_spatial} can be reinterpreted as a penalised weighted least squares (WLS) problem, which clarifies what the spatial update computes and corresponds to the system actually solved in the implementation. Introduce the per-observation Newton target $r_i = -g_i/h_i$, which represents the update in $F_i$ that would drive the gradient at observation $i$ to zero if the loss were perfectly quadratic. We then examine the WLS problem that determines a spatial update $\Delta\mathbf{z}$ by minimising the weighted deviation from these targets, subject to a smoothness penalty on the resulting field:
\begin{equation}\label{eq:wls}
\min_{\Delta\mathbf{z}}\; \frac{1}{2}\sum_{i=1}^{N} h_i\bigl([\mathbf{A}\Delta\mathbf{z}]_i - r_i\bigr)^2 + \frac{\lambda_{\mathrm{spa}}}{2}\,(\mathbf{z} + \Delta\mathbf{z})^\top\mathbf{Q}\,(\mathbf{z} + \Delta\mathbf{z})\,.
\end{equation}
By expanding the penalty term and taking the derivative with respect to $\Delta\mathbf{z}$, we obtain the normal equations corresponding to~\eqref{eq:wls}: $(\mathbf{A}^\top\mathbf{H}\mathbf{A} + \lambda_{\mathrm{spa}}\mathbf{Q})\,\Delta\mathbf{z} = \mathbf{A}^\top\mathbf{H}\mathbf{r} - \lambda_{\mathrm{spa}}\mathbf{Q}\mathbf{z}$. Since $(\mathbf{H}\mathbf{r})_i = h_i \cdot (-g_i/h_i) = -g_i$, the right-hand side simplifies to $-(\mathbf{A}^\top\mathbf{g} + \lambda_{\mathrm{spa}}\mathbf{Q}\mathbf{z})$, which exactly reproduces the Newton system~\eqref{eq:newton_spatial}. The spatial solver therefore computes the mesh coefficients that best approximate the per-observation Newton targets $r_i$ in a weighted least-squares sense, where each observation's weight $h_i$ reflects the local curvature of the loss. Under logistic loss, for instance, $h_i = \hat{p}_i(1 - \hat{p}_i)$, so observations with uncertain predictions receive the most influence. The GMRF penalty ensures that the resulting field remains spatially smooth.

Both stages originate from the same second-order expansion of the penalised objective, and both fit the same per-observation Newton targets $r_i$ with weights $h_i$. Given the tree structure, Stage~1 fits them with a constant in each leaf, whereas Stage~2 fits them with a smooth field on the mesh, regularised by the GMRF penalty.  Section~\ref{sec:algorithm} combines these updates into the complete boosting algorithm.

\subsection{The two-stage boosting algorithm}\label{sec:algorithm}

The complete algorithm repeatedly applies the two Newton steps introduced in Section~\ref{sec:objective} inside a boosting loop. At iteration $t$, given the current predictor $F^{(t-1)}$, the update proceeds as follows.

\textbf{Stage~1: covariate Newton step (tree update).} First, compute the gradients $g_i$ and Hessians $h_i$ based on $F^{(t-1)}$. Next, fit a regression tree $f_t$ to the negative gradient $-\mathbf{g}$ using only the covariate matrix $\mathbf{X}$. Determine the Newton leaf values $\hat{c}_k$ using~\eqref{eq:newton_leaf}. Finally, update the covariate part and the overall predictor as
\begin{equation*}
F_{\mathrm{cov}}^{(t)} = F_{\mathrm{cov}}^{(t-1)} + \eta\,f_t(x_i)\,, \qquad F^{(t,\mathrm{mid})} = F^{(t-1)} + \eta\,f_t(x_i)\,.
\end{equation*}

\textbf{Stage~2: spatial Newton step (mesh update).} Re-evaluate the gradient and Hessian at the mid-round predictor $F^{(t,\mathrm{mid})}$. Solve the regularised Newton system~\eqref{eq:newton_spatial} to obtain $\Delta\mathbf{z}^{(t)}$. Apply the learning rate $\eta$:
\begin{equation}\label{eq:spatial_update}
\mathbf{z}^{(t)} = \mathbf{z}^{(t-1)} + \eta\,\Delta\mathbf{z}^{(t)}\,.
\end{equation}
Then update the predictor via $F^{(t)} = F^{(t,\mathrm{mid})} + \eta\,\mathbf{A}\,\Delta\mathbf{z}^{(t)}$.

The procedure is described in Algorithm~\ref{alg:inlaboost}. The algorithm outputs three components: the covariate ensemble $F_{\mathrm{cov}}$, which is used to interpret individual covariate effects; the spatial field $\mathbf{z}$, which captures the geographic component; and the combined predictor $F$, which is used for prediction on new data. Both stages use the same learning rate $\eta$, selected by cross-validation. Separate rates for the tree and spatial updates are possible, but in our experiments they added a tuning parameter without a noticeable gain in performance. The mesh geometry ($M$, $\mathbf{A}$, $\mathbf{Q}$) and the SPDE range parameter $\kappa$ are fixed before boosting and do not change across iterations. Because each candidate value of $\kappa$ produces a different precision matrix, $\kappa$ is included in the cross-validation grid alongside the algorithmic parameters $\eta$, $\lambda_{\mathrm{spa}}$, $\lambda_{\mathrm{tree}}$, $T$, tree depth, and minimum samples per leaf.

\begin{algorithm}[htbp]
\caption{SAMBoost}\label{alg:inlaboost}
\begin{algorithmic}[1]
\Require Data $\{(y_i, x_i, s_i)\}_{i=1}^N$; mesh with $M$ nodes, projection matrix $\mathbf{A}$, precision matrix $\mathbf{Q}$; learning rate $\eta$; penalties $\lambda_{\mathrm{tree}}$, $\lambda_{\mathrm{spa}}$; number of rounds $T$.
\Statex
\State Initialise $F_0 \leftarrow \arg\min_{\beta_0} \sum_{i=1}^{N} \ell(y_i, \beta_0)$, $\;F \leftarrow F_0\,\mathbf{1}_N$, $\;F_{\mathrm{cov}} \leftarrow \mathbf{0}$, $\;\mathbf{z} \leftarrow \mathbf{0}$ 
\Statex
\For{$t = 1, \ldots, T$}
    \Statex \hspace{\algorithmicindent}\textit{// Stage 1: covariate Newton step}
    \State $g_i \leftarrow \partial\ell/\partial F\big|_{F_i}$, $\;h_i \leftarrow \partial^2\ell/\partial F^2\big|_{F_i}$ \hfill for $i = 1,\ldots,N$
    \State Fit tree $f_t$ to $-\mathbf{g}$ using features $\mathbf{X}$ only
    \State Compute Newton leaf values $\hat{c}_{k} \leftarrow -G_k/(H_k + \lambda_{\mathrm{tree}})$
    \State $F_{\mathrm{cov}} \leftarrow F_{\mathrm{cov}} + \eta\,f_t(\mathbf{X})$, $\;F \leftarrow F + \eta\,f_t(\mathbf{X})$
    \Statex
    \Statex \hspace{\algorithmicindent}\textit{// Stage 2: spatial Newton step}
    \State $g_i \leftarrow \partial\ell/\partial F\big|_{F_i}$, $\;h_i \leftarrow \partial^2\ell/\partial F^2\big|_{F_i}$ \hfill for $i = 1,\ldots,N$
    \State Solve $(\mathbf{A}^\top\mathbf{H}\mathbf{A} + \lambda_{\mathrm{spa}}\,\mathbf{Q})\,\Delta\mathbf{z} = -(\mathbf{A}^\top\mathbf{g} + \lambda_{\mathrm{spa}}\,\mathbf{Q}\mathbf{z})$
    \State $\mathbf{z} \leftarrow \mathbf{z} + \eta\,\Delta\mathbf{z}$, $\;F \leftarrow F + \eta\,\mathbf{A}\,\Delta\mathbf{z}$
\EndFor
\Statex
\State \textbf{return} $F_{\mathrm{cov}}$, $\mathbf{z}$, $F$
\end{algorithmic}
\end{algorithm}

\section{Simulation study}\label{sec:simulations}

\subsection{Design}\label{sec:sim-design}

To evaluate SAMBoost and benchmark it against the two baselines introduced in Section~\ref{sec:contributions}, we construct two simulation experiments with known ground truth. In both setups, the response is generated from a linear predictor that is split additively into a covariate term $F^{*}_{\mathrm{cov}}(x)$, depending solely on observed features, and a spatial term $F^{*}_{\mathrm{spa}}(s)$, depending solely on geographic location. The asterisk distinguishes these true components from the model components of Section~\ref{sec:model}, whose estimates are denoted $\hat{F}_{\mathrm{cov}}$ and $\hat{F}_{\mathrm{spa}}$. Both components are specified in closed form, so our assessment covers not only the overall predictive performance but also the extent to which each method recovers each component separately. Spatial positions $s_i$ are sampled uniformly on the unit square $[0,1]^2$, with $N = 1000$ training and 1000 test observations drawn independently. Each experiment is replicated with 30 independent random seeds and all metrics are summarised as mean $\pm$ standard deviation across these replications. Statistical significance is evaluated using paired Wilcoxon signed-rank tests~\citep{Wilcoxon1945}, a nonparametric procedure that tests whether seed-level metric differences between two methods systematically deviate from zero, based on the 30 repeated runs. The simulations are designed as follows. 

\begin{figure}[htbp]
\centering
\begin{subfigure}[t]{\textwidth}
\centering
\includegraphics[width=\textwidth]{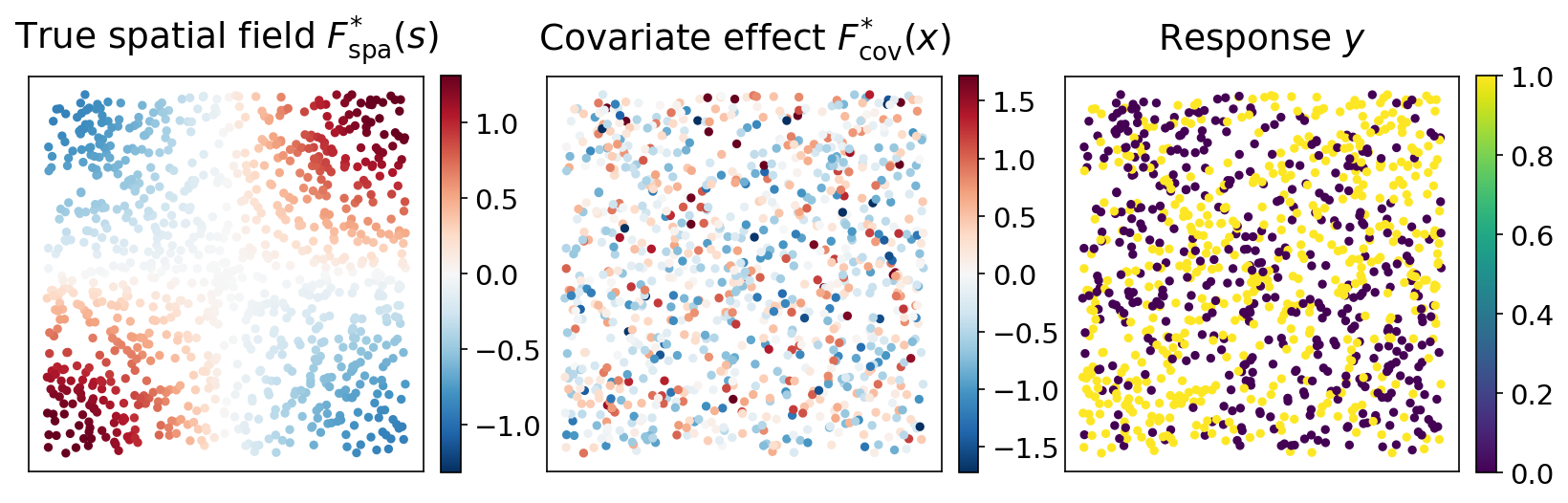}
\caption{Simulation 1: chequerboard spatial field $F^{*}_{\mathrm{spa}}(s)$, covariate effect $F^{*}_{\mathrm{cov}}(x)$ with no spatial structure, and binary response $y$.}\label{fig:sim1-domain}
\end{subfigure}

\vspace{0.5em}

\begin{subfigure}[t]{\textwidth}
\centering
\includegraphics[width=\textwidth]{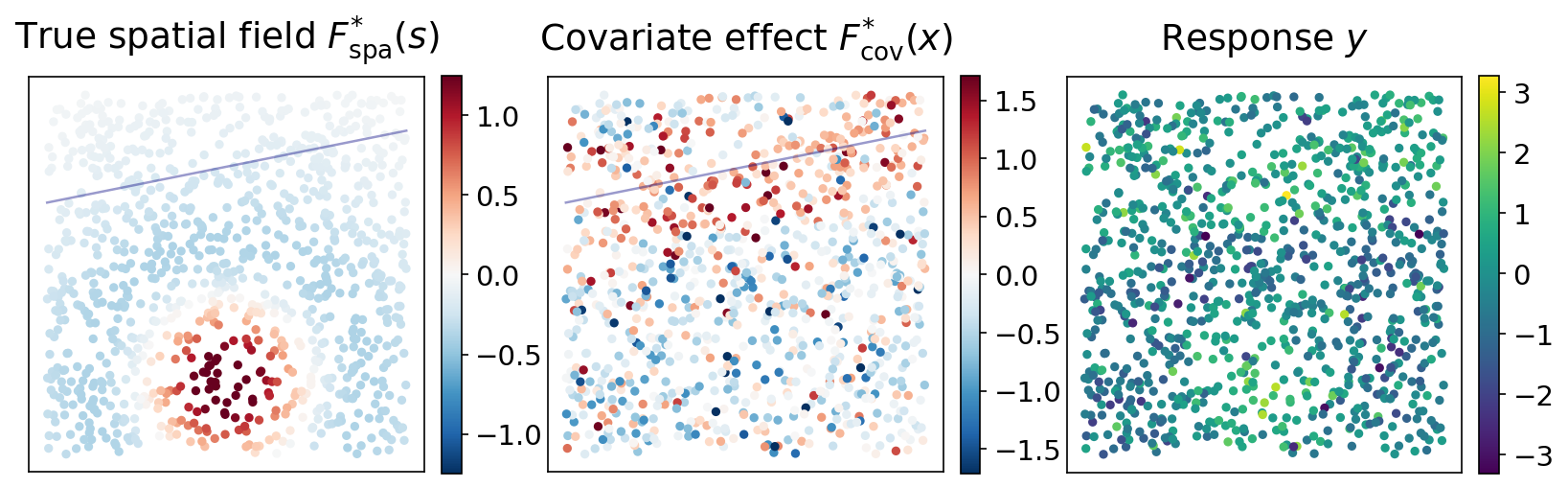}
\caption{Simulation 2: true spatial field $F^{*}_{\mathrm{spa}}(s)$ with a single hotspot in the lower centre, covariate effect $F^{*}_{\mathrm{cov}}(x)$ with a faint positive band along the river corridor (diagonal line), and continuous response $y$.}\label{fig:sim2-domain}
\end{subfigure}
\caption{Simulation domains and data-generating processes.}\label{fig:sim-domains}
\end{figure}

\medskip
\textbf{Simulation 1: classification without confounding.}
The first experiment assesses the methods on a classification task in the absence of spatial confounding. The response follows a Bernoulli distribution with a logit link
\begin{equation}\label{eq:sim1-response}
y_i \sim \mathrm{Bernoulli}\bigl(\sigma(F^{*}_{\mathrm{cov}}(x_i) + F^{*}_{\mathrm{spa}}(s_i))\bigr)\,,
\end{equation}
where $\sigma$ is the logistic function and the corresponding loss is the log loss. The spatial effect is a chequerboard pattern
\begin{equation}\label{eq:sim1-checker}
F^{*}_{\mathrm{spa}}(s) = 1.4\,\cos(\pi s_1)\cos(\pi s_2)\,,
\end{equation}
yielding alternating positive and negative quadrants. The negative lobes are then multiplied by $0.65$, so that the field is asymmetric. The covariate effect is a simple, almost additive function
\begin{equation}\label{eq:sim1-fcov}
F^{*}_{\mathrm{cov}}(x) = 1.5\,x_1 + x_2^2 + 3\,\mathbf{1}\{x_3 > 0\} + x_4\,x_5\,,
\end{equation}
where covariates $x_1, \ldots, x_5$ are sampled independently from $\mathcal{N}(0,1)$, and two additional noise covariates $x_{\mathrm{noise},1}$ and $x_{\mathrm{noise},2}$ are added to the design matrix but excluded from $F^{*}_{\mathrm{cov}}$. The function $F^{*}_{\mathrm{cov}}$ is centred and scaled to satisfy $\mathrm{Var}(F^{*}_{\mathrm{cov}}) = 0.5$. The spatial field adds a further variance of about $0.35$, giving a total signal variance of approximately $0.85$. In logistic regression the implicit noise variance is $\pi^2/3 \approx 3.29$ (the variance of a standard logistic distribution \citep{johnson1994continuous}), so the signal-to-noise ratio is roughly $0.26$, producing a moderately difficult classification task. Figure~\ref{fig:sim1-domain} shows the chequerboard field, the covariate effect, and the binary response for Simulation~1.

\medskip
\textbf{Simulation 2: regression with a spatially confounded covariate.}
The second simulation introduces spatial confounding via a covariate that is fully determined by spatial location. The response is generated from a Gaussian regression model with identity link,
\begin{equation}\label{eq:sim2-response}
y_i = F^{*}_{\mathrm{cov}}(x_i) + F^{*}_{\mathrm{spa}}(s_i) + \varepsilon_i\,, \qquad \varepsilon_i \sim \mathcal{N}(0,\, 0.5^2)\,.
\end{equation}
The spatial effect consists of a single Mexican-hat hotspot centred at $c = (0.5, 0.2)$,
\begin{equation}\label{eq:sim2-hotspot}
F^{*}_{\mathrm{spa}}(s) = 2.5\,\Bigl[\exp\!\Bigl(-\frac{\|s - c\|^2}{2 \times 0.15^2}\Bigr) - 0.35\,\exp\!\Bigl(-\frac{\|s - c\|^2}{2 \times 0.35^2}\Bigr)\Bigr]\,,
\end{equation}
which yields a region of elevated risk at the centre, surrounded by a ring of reduced risk, with the effect decaying back toward zero away from the hotspot.

The covariate contribution combines linear, threshold, and interaction terms,
\begin{equation}\label{eq:sim2-fcov}
F^{*}_{\mathrm{cov}}(x) = 1.5\,x_1 + 3\,\mathbf{1}\{x_3 > 0\} + 2.5\,x_1 x_2 + 2\,x_2 x_3 + 1.5\,x_4 x_5 + 4.5\,x_{\mathrm{river}}\,.
\end{equation}
Covariates $x_1, \ldots, x_5$ are drawn independently from $\mathcal{N}(0,1)$. Two additional noise covariates, $x_{\mathrm{noise},1}$ and $x_{\mathrm{noise},2}$, are appended to the feature matrix but excluded from $F^{*}_{\mathrm{cov}}$, providing a check on each model’s ability to discard irrelevant predictors. The covariate $x_{\mathrm{river}}$ is a binary indicator set to one within a corridor of half-width $D = 0.15$ around a diagonal line segment from $(0, 0.70)$ to $(1, 0.90)$, and zero otherwise. Because $x_{\mathrm{river}}$ is completely determined by spatial position, it induces confounding between the covariate and spatial components, forcing the model to partition the predictive signal at each location between the observed feature and the latent spatial risk.

The function $F^{*}_{\mathrm{cov}}$ is centred and rescaled so that $\mathrm{Var}(F^{*}_{\mathrm{cov}}) = 0.5$. The spatial field adds roughly $0.16$ of variance and the observation noise has variance $0.25$, leading to an overall signal-to-noise ratio of $(\mathrm{Var}(F^{*}_{\mathrm{cov}}) + \mathrm{Var}(F^{*}_{\mathrm{spa}}))/\mathrm{Var}(\varepsilon) \approx 2.6$. By construction, the covariate effect is stronger than the spatial effect ($\mathrm{Var}(F^{*}_{\mathrm{cov}})/\mathrm{Var}(F^{*}_{\mathrm{spa}}) \approx 3.2$), since this setting is intended to challenge covariate recovery in the presence of confounding. Figure~\ref{fig:sim2-domain} shows the hotspot, the covariate effect with the river corridor visible as a faint positive band from the $4.5\,x_{\mathrm{river}}$ term, and the continuous response for Simulation~2.

\subsection{Baselines and evaluation}\label{sec:sim-baselines}

We compare two baseline models to SAMBoost: GPBoost and INLA.

\textbf{GPBoost} \citep{Sigrist2022,Sigrist2023} (v1.6.6) fits a tree ensemble and a Gaussian process random effect jointly within a single likelihood, using an exponential covariance function. The GP covariance hyperparameters (range and variance) are estimated internally via marginal likelihood optimisation and are not included in the tuning grid. We considered both the exact GP and the Vecchia approximation, treating the number of Vecchia neighbours as a tuning parameter. In the simulation experiments, GPBoost employs exact GP inference, which is feasible at $N = 1000$, whereas the scaling analysis in Section~\ref{sec:scaling} relies on the Vecchia approximation, which becomes necessary at larger sample sizes. The tree ensemble models the covariate effect, while the spatial effect used for dissociation assessment is given by the posterior kriging prediction of the GP random effect at the test locations, conditional on the fitted trees.

\textbf{INLA} \citep{RueMartinoCHopin2009,LindgrenRue2015} (R-INLA v24.12.11) fits a Bayesian spatial model that combines an SPDE Gaussian random field with covariate effects. Discrete covariates are included as linear fixed effects, while continuous covariates are represented by smooth nonlinear functions constructed from one-dimensional SPDE splines with penalised complexity (PC) Mat\'ern priors and a sum-to-zero constraint. The spatial prior’s SPDE range is initialised using an empirical variogram estimated from the training data. Inference relies on the empirical Bayes integration strategy, which cuts computation time by about 25\% with negligible influence on the evaluation metrics. The covariate component is defined as the sum of all estimated covariate effects, and the spatial component used for dissociation analysis is the posterior mean of the SPDE field mapped to the test locations.

\medskip
The prediction quality is measured on the test set, with $\hat{p}_i$ and $\hat{y}_i$ denoting respectively the predicted probability and the predicted response for observation $i$. For the classification experiment we report AUC (area under the ROC curve) \citep{fawcett2006introduction}, log loss $= -N^{-1}\sum_i[y_i\log\hat{p}_i + (1-y_i)\log(1-\hat{p}_i)]$, and Brier score $= N^{-1}\sum_i(\hat{p}_i - y_i)^2$. For the regression we report RMSE $= (N^{-1}\sum_i(\hat{y}_i - y_i)^2)^{1/2}$.
We assess dissociation quality by comparing the recovered components to the known ground truth. For each method, we compute Pearson’s correlation coefficient ($r$) and the RMSE between $\hat{F}_{\mathrm{cov}}$ and the true $F^{*}_{\mathrm{cov}}(x)$, and between $\hat{F}_{\mathrm{spa}}$ and the true $F^{*}_{\mathrm{spa}}(s)$. For the spatial component, correlation and RMSE can yield different rankings because correlation measures whether the shape of the spatial field is correctly reproduced, regardless of its amplitude, whereas RMSE also penalises errors in magnitude. A method that captures the spatial pattern faithfully but shrinks its peaks toward zero may obtain a high correlation but a poor RMSE. We therefore report both metrics and treat RMSE as the more reliable indicator of field recovery.

\medskip
Each method was tuned according to its own recommended procedure, and cross-validation used identical folds across methods to ensure fair comparison. The two tree-based approaches (SAMBoost and GPBoost) were optimised via grid search over learning rate, tree depth, minimum samples per leaf, and regularisation parameters, using 3-fold stratified cross-validation on the training set. Model selection was based on AUC for Simulation~1 and RMSE for Simulation~2. For SAMBoost, the SPDE range parameter $\kappa$ was additionally included in the search grid, since it sets the spatial range and directly affects predictive performance. GPBoost determined the number of boosting iterations automatically using early stopping within \texttt{gpb{.}cv}, the package’s built-in tuning routine, whereas SAMBoost treated the number of rounds as a grid search parameter. For INLA, we tuned the PC prior hyperparameters (prior range and prior standard deviation) via grid search, choosing the setting with the highest Conditional Predictive Ordinate (CPO). CPO is the standard model comparison metric in INLA, and implementing AUC in INLA’s leave-one-out scheme is not straightforward. The grids contain up to 50 candidate configurations, except for INLA, which uses 9. This smaller grid reflects the fact that INLA estimates its model hyperparameters internally through marginal likelihood, so only the PC prior parameters remain for external tuning.

All experiments were run on a 12-core Intel CPU (2.10 GHz) with 64 GB of RAM. All reported times are wall-clock fit times. We used Python~3.12 for all methods, except for INLA, which was run in R~4.4.3. Both SAMBoost and INLA rely on SPDE meshes with roughly 900 nodes, featuring a finer resolution in the interior of the domain and a coarser extension near the outer boundary.    The complete source code, tuning grids, and experimental pipeline are publicly available.\footnote{\url{https://github.com/MoriahMulah/SAMBoost}}

\subsection{Classification results}\label{sec:sim1-results}
Table~\ref{tab:sim1-results} presents the Simulation~1 outcomes across 30 random seeds. Without confounding, the spatial methods exhibit only minor differences in predictive performance. INLA attains the highest AUC ($0.699 \pm 0.019$), with SAMBoost ($0.689 \pm 0.020$) and GPBoost ($0.678 \pm 0.021$) close behind.

\begin{table}[htbp]
\centering
\caption{Simulation 1 (classification). Prediction and dissociation metrics on the test set, reported as mean $\pm$ standard deviation over 30 seeds. Best value per column in bold.}\label{tab:sim1-results}
\smallskip
\small
\resizebox{\textwidth}{!}{%
\begin{tabular}{l
  S[table-format=1.3] @{${}\pm{}$} S[table-format=1.3]
  S[table-format=1.3] @{${}\pm{}$} S[table-format=1.3]
  S[table-format=1.3] @{${}\pm{}$} S[table-format=1.3]
  S[table-format=1.3] @{${}\pm{}$} S[table-format=1.3]
  S[table-format=1.3] @{${}\pm{}$} S[table-format=1.3]
  S[table-format=1.3] @{${}\pm{}$} S[table-format=1.3]
  S[table-format=1.3] @{${}\pm{}$} S[table-format=1.3]
  S[table-format=2.1] @{${}\pm{}$} S[table-format=2.1]}
\toprule
& \multicolumn{2}{c}{AUC $\uparrow$}
& \multicolumn{2}{c}{Log loss $\downarrow$}
& \multicolumn{2}{c}{Brier $\downarrow$}
& \multicolumn{2}{c}{$r_{\mathrm{spa}}$ $\uparrow$}
& \multicolumn{2}{c}{$\mathrm{rmse}_{\mathrm{spa}}$ $\downarrow$}
& \multicolumn{2}{c}{$r_{\mathrm{cov}}$ $\uparrow$}
& \multicolumn{2}{c}{$\mathrm{rmse}_{\mathrm{cov}}$ $\downarrow$}
& \multicolumn{2}{c}{Time (s)} \\
\midrule
SAMBoost     & 0.689 & 0.020 & 0.637 & 0.012 & 0.223 & 0.005 & \bfseries 0.951 & 0.021 & \bfseries 0.221 & 0.035 & 0.749 & 0.040 & \bfseries 0.476 & 0.033 & \bfseries 0.8 & 0.4 \\
GPBoost      & 0.678 & 0.021 & 0.645 & 0.011 & 0.227 & 0.005 & 0.926 & 0.027 & 0.268 & 0.037 & 0.707 & 0.057 & 0.557 & 0.065 & 80.7 & 49.7 \\
INLA         & \bfseries 0.699 & 0.019 & \bfseries 0.629 & 0.011 & \bfseries 0.220 & 0.005 & \bfseries 0.951 & 0.020 & 0.281 & 0.082 & \bfseries 0.829 & 0.027 & 0.528 & 0.083 & 40.8 & 39.5 \\
\bottomrule
\end{tabular}}
\end{table}

Dissociation metrics provide only limited additional differentiation. SAMBoost attains the most accurate spatial field reconstruction ($\mathrm{rmse}_{\mathrm{spa}} = 0.221$) and the best recovery of covariate magnitudes ($\mathrm{rmse}_{\mathrm{cov}} = 0.476$). INLA yields the highest covariate correlation ($r_{\mathrm{cov}} = 0.829$), due to its correctly specified spline model that reflects the additive structure of $F^{*}_{\mathrm{cov}}$. However, its spatial RMSE is higher ($0.281$) despite matching SAMBoost on spatial correlation ($r_{\mathrm{spa}} = 0.951$), confirming that the INLA field reproduces the chequerboard shape faithfully, but underestimates its amplitude. GPBoost reconstructs the spatial pattern less accurately than both SAMBoost and INLA ($r_{\mathrm{spa}} = 0.926$). These differences in dissociation are relatively small; the following simulation, which incorporates interactions and spatial confounding, shows where the methods diverge substantially.

\begin{figure}[htbp]
\centering
\includegraphics[width=\textwidth]{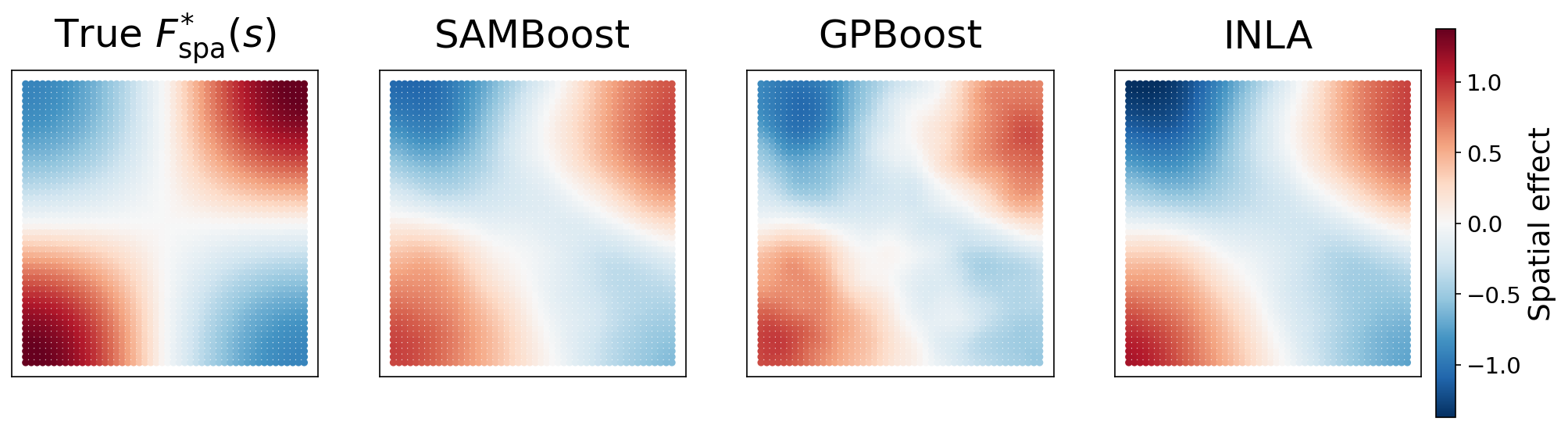}
\caption{Recovered spatial fields in Simulation 1 for a representative seed, namely the seed for which the spatial RMSE of SAMBoost is the median over the 30 replications.}\label{fig:sim1-fields}
\end{figure}

Figure~\ref{fig:sim1-fields} presents the reconstructed spatial fields for a representative seed, confirming that SAMBoost preserves the chequerboard amplitudes more faithfully than INLA and GPBoost, which both attenuate the peaks. SAMBoost is also the fastest method, fitting in under 1 second compared to 41 seconds for INLA and 81 seconds for GPBoost (Table~\ref{tab:sim1-results}).

\subsection{Regression results}\label{sec:sim2-results}

Table~\ref{tab:sim2-results} presents the results of Simulation~2 across 30 random seeds, a setting in which spatial confounding and covariate interactions together pose difficulties for the methods. Two clear patterns can be observed.

\begin{table}[htbp]
\centering
\caption{Simulation 2 (regression with confounding). Prediction and dissociation metrics on the test set, reported as mean $\pm$ standard deviation over 30 seeds. Best value per column in bold.}\label{tab:sim2-results}
\smallskip
\small
\resizebox{\textwidth}{!}{%
\begin{tabular}{l
  S[table-format=1.3] @{${}\pm{}$} S[table-format=1.3]
  S[table-format=1.3] @{${}\pm{}$} S[table-format=1.3]
  S[table-format=1.3] @{${}\pm{}$} S[table-format=1.3]
  S[table-format=1.3] @{${}\pm{}$} S[table-format=1.3]
  S[table-format=1.3] @{${}\pm{}$} S[table-format=1.3]
  S[table-format=3.1] @{${}\pm{}$} S[table-format=2.1]}
\toprule
& \multicolumn{2}{c}{RMSE $\downarrow$}
& \multicolumn{2}{c}{$\mathrm{rmse}_{\mathrm{spa}}$ $\downarrow$}
& \multicolumn{2}{c}{$r_{\mathrm{spa}}$ $\uparrow$}
& \multicolumn{2}{c}{$r_{\mathrm{cov}}$ $\uparrow$}
& \multicolumn{2}{c}{$\mathrm{rmse}_{\mathrm{cov}}$ $\downarrow$}
& \multicolumn{2}{c}{Time (s)} \\
\midrule
SAMBoost     & \bfseries 0.596 & 0.020 & \bfseries 0.160 & 0.018 & 0.938 & 0.020 & \bfseries 0.895 & 0.013 & \bfseries 0.318 & 0.020 & \bfseries 6.6 & 3.1 \\
GPBoost      & 0.610 & 0.020 & 0.207 & 0.034 & 0.888 & 0.042 & 0.866 & 0.030 & 0.371 & 0.032 & 111.9 & 26.0 \\
INLA         & 0.754 & 0.024 & 0.207 & 0.069 & \bfseries 0.947 & 0.016 & 0.625 & 0.020 & 0.593 & 0.032 & 72.5 & 25.7 \\
\bottomrule
\end{tabular}}
\end{table}

First, SAMBoost achieves the strongest performance ($\mathrm{RMSE} = 0.596 \pm 0.020$), surpassing GPBoost ($0.610 \pm 0.020$). Both methods outperform INLA by a wide margin ($0.754 \pm 0.024$). The advantage of the tree-based approaches over INLA is primarily due to better covariate recovery. Since the additive model cannot capture the interaction terms in $F^{*}_{\mathrm{cov}}$, the residual covariate structure is absorbed into the spatial field, which harms both components. SAMBoost outperforms INLA on all 30 seeds for $r_{\mathrm{cov}}$ and $\mathrm{rmse}_{\mathrm{cov}}$ (paired Wilcoxon signed-rank test, $p < 10^{-4}$).

Second, SAMBoost provides stronger dissociation than GPBoost. SAMBoost performs better on $\mathrm{rmse}_{\mathrm{spa}}$ for 29 of 30 seeds ($p < 10^{-4}$) and on $\mathrm{rmse}_{\mathrm{cov}}$ for 22 of 30 seeds ($p = 0.0004$). This underscores a key aspect of spatial modelling: improved overall prediction does not guarantee a good decomposition, since one component may capture a signal that properly belongs to the other. Dissociation metrics expose this kind of misattribution, which remains invisible if one looks only at the overall RMSE. The two spatial metrics reveal an instructive discrepancy. INLA attains the strongest spatial correlation ($r_{\mathrm{spa}} = 0.947$), exceeding both SAMBoost ($0.938$) and GPBoost ($0.888$). However, its spatial RMSE ($0.207$) is noticeably larger than that of SAMBoost ($0.160$). This mismatch arises because INLA treats the river term as a linear fixed effect, which reduces confounding leakage and preserves the spatial field’s structure (yielding high correlation), but the purely additive covariate specification cannot capture the interaction residual, thereby increasing the spatial magnitude error. 

\begin{figure}[htbp]
\centering
\includegraphics[width=\textwidth]{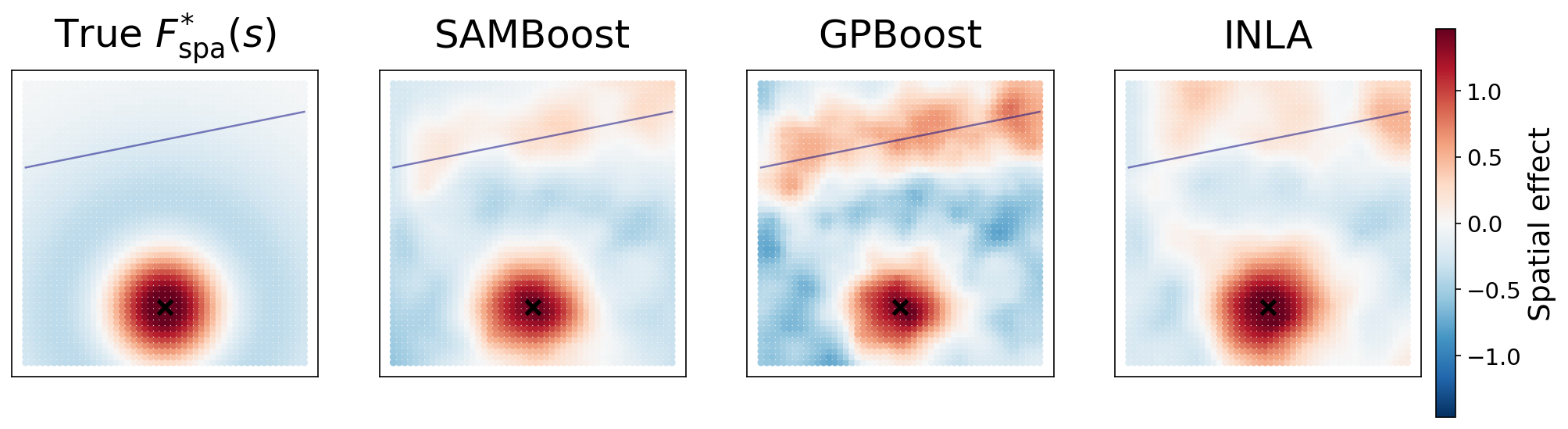}
\caption{Recovered spatial fields in Simulation 2 for a representative seed, namely the seed for which the spatial RMSE of SAMBoost is the median over the 30 replications. The cross marks the hotspot centre and the diagonal line marks the river corridor.}\label{fig:sim2-fields}
\end{figure}

Figure~\ref{fig:sim2-fields} presents the recovered spatial fields for a representative seed. SAMBoost cleanly reconstructs the hotspot, achieving the lowest RMSE and showing only limited background fluctuations. GPBoost also identifies the hotspot, but introduces false positive patches in the upper area, where the confounding river corridor contaminates its spatial surface. INLA limits this leakage through its linear river term, but its field is affected by the unmodelled interaction residual discussed above. SAMBoost fits in about 7 seconds, compared with 73 seconds for INLA and 112 seconds for GPBoost. Differences in fit time across simulations are mainly due to the hyperparameters chosen by cross-validation, in particular the number of boosting rounds selected for SAMBoost (200 in Simulation~2 versus 30 in Simulation~1) given the more complex covariate structure.

\subsection{Role of covariate complexity}\label{sec:ablation}

The two simulations combine multiple differences (response type, spatial field, and confounding), which makes it hard to pinpoint why SAMBoost performs better. To specifically assess the impact of covariate interactions, we conduct a sensitivity analysis that retains the spatial layout, river confounder, and regression setup of Simulation~2, but replaces the interaction-rich covariate function~\eqref{eq:sim2-fcov} with a simpler, nearly additive specification in which the three interaction terms ($x_1 x_2$, $x_2 x_3$, $x_4 x_5$) are reduced to a single weaker interaction ($x_4 x_5$) and a quadratic term ($x_2^2$),
\begin{equation}\label{eq:fsimple}
F^{*}_{\mathrm{cov,simple}}(x) = 1.5\,x_1 + x_2^2 + 3\,\mathbf{1}\{x_3 > 0\} + x_4\,x_5 + 4.5\,x_{\mathrm{river}}\,.
\end{equation} Relative to the interaction-rich $F^{*}_{\mathrm{cov}}$ in Simulation~2, this simpler function is easier to capture with an additive model without interactions. Indeed, fitting such a model to $F^{*}_{\mathrm{cov}}$ directly yields an $R^2$ of about $0.25$ for the interaction-rich version, but above $0.85$ for $F^{*}_{\mathrm{cov,simple}}$, showing that covariate interactions are the main challenge for additive approaches. Table~\ref{tab:ablation} presents the dissociation metrics for SAMBoost, INLA and GPBoost under both covariate setups.

\begin{table}[htbp]
\centering
\caption{Sensitivity analysis on covariate complexity. Dissociation metrics under a simple (nearly additive) and interaction-rich covariate function, with the same spatial design and river confounder. Reported as 30-seed means. Best value per column within each covariate setting in bold.}\label{tab:ablation}
\smallskip
\small
\begin{tabular}{ll
  S[table-format=1.3]
  S[table-format=1.3]
  S[table-format=1.3]
  S[table-format=1.3]}
\toprule
& & \multicolumn{2}{c}{Covariate recovery}
& \multicolumn{2}{c}{Spatial recovery} \\
\cmidrule(lr){3-4} \cmidrule(lr){5-6}
{Covariate function} & {Method} & {$r_{\mathrm{cov}}$ $\uparrow$} & {$\mathrm{rmse}_{\mathrm{cov}}$ $\downarrow$} & {$r_{\mathrm{spa}}$ $\uparrow$} & {$\mathrm{rmse}_{\mathrm{spa}}$ $\downarrow$} \\
\midrule
Simple $F^{*}_{\mathrm{cov,simple}}$           & SAMBoost & \bfseries 0.931 & \bfseries 0.259 & 0.939 & \bfseries 0.159 \\
                     & INLA      & 0.924 & 0.283 & \bfseries 0.962 & 0.194 \\
                     & GPBoost   & 0.912 & 0.305 & 0.889 & 0.207 \\
\addlinespace
Interaction-rich $F^{*}_{\mathrm{cov}}$ & SAMBoost & \bfseries 0.895 & \bfseries 0.318 & 0.938 & \bfseries 0.160 \\
                     & INLA      & 0.625 & 0.593 & \bfseries 0.947 & 0.207 \\
                     & GPBoost   & 0.866 & 0.371 & 0.888 & 0.207 \\
\bottomrule
\end{tabular}
\end{table}

With the simple covariate function, INLA attains nearly the same covariate recovery as SAMBoost (Table~\ref{tab:ablation}), showing that a correctly specified additive model performs well when the true covariate effect is close to additive. When the covariate function is rich in interactions, however, INLA’s covariate recovery deteriorates sharply ($r_{\mathrm{cov}}$ falls from $0.924$ to $0.625$), whereas SAMBoost remains stable ($0.931$ to $0.895$). This leakage extends beyond the covariate component: INLA’s spatial RMSE worsens from $0.194$ to $0.207$, implying that the unmodelled interaction residual also contaminates the spatial field.
SAMBoost’s spatial recovery remains effectively unchanged as covariate complexity increases ($\mathrm{rmse}_{\mathrm{spa}}$: $0.159$ to $0.160$). GPBoost exhibits a similar degree of spatial robustness ($\mathrm{rmse}_{\mathrm{spa}}$: $0.207$ in both settings), confirming that the collapse is specific to the additive covariate specification rather than to joint spatial estimation in general. Nevertheless, GPBoost’s spatial field consistently absorbs the river confounder ($\mathrm{rmse}_{\mathrm{spa}}$: $0.207$ compared with $0.160$ for SAMBoost), meaning that only SAMBoost accurately recovers both components under the interaction-rich setup.
This sensitivity analysis isolates the two benefits: relative to GPBoost, the spatial gain is inherent and persistent; relative to INLA, the covariate benefit arises only when the true covariate function includes interactions that go beyond what an additive model can capture.

\subsection{Computational scaling}\label{sec:scaling}

To examine how training time scales with sample size $N$, we record wall-clock time as a function of $N$ while keeping all other settings constant. A single SPDE mesh (about 1,650 nodes) is constructed once and reused for every value of $N$, all boosting approaches employ 100 trees, and each setup is fitted sequentially (one model at a time) to prevent CPU contention. All methods use the same datasets. $N$ ranges from 500 to 64,000 across three random seeds, and the recorded times are averaged.

\begin{figure}[htbp]
\centering
\includegraphics[width=0.75\textwidth]{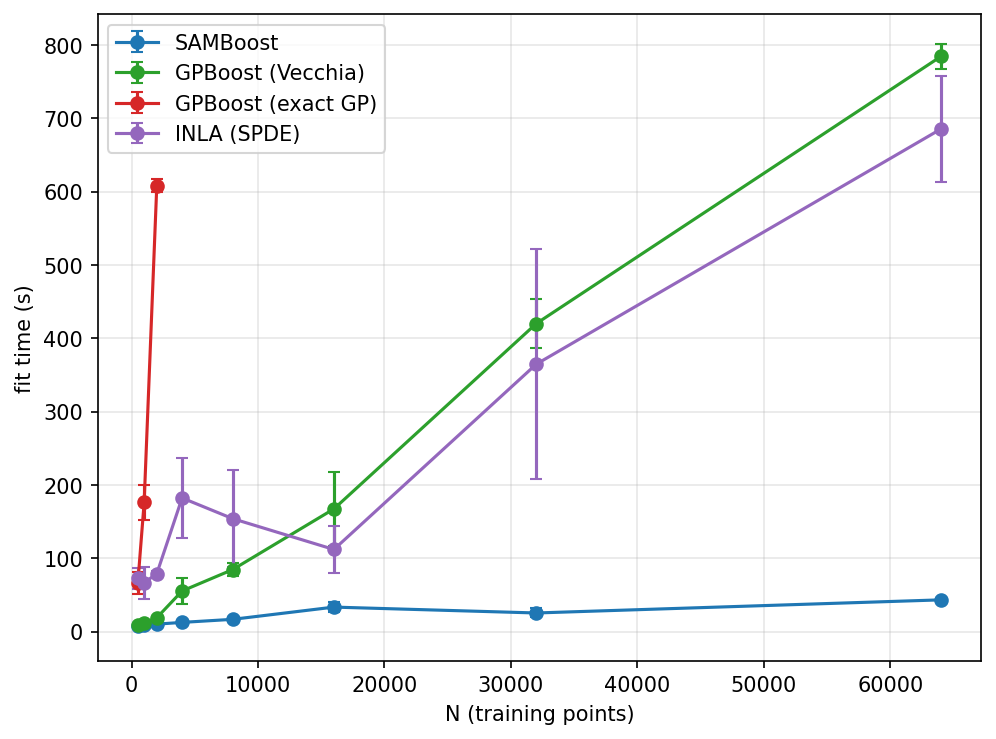}
\caption{Fit time as a function of the number of training points $N$, with a fixed mesh and 100 trees. Points show the mean over three seeds and error bars the mean plus or minus the standard deviation.}\label{fig:scaling}
\end{figure}

Figure~\ref{fig:scaling} presents the results on linear axes. SAMBoost is the fastest method for all $N$ and also exhibits the most favourable scaling, with an empirical complexity of about $N^{0.37}$. This sublinear scaling occurs because the main computational cost is solving the SPDE system on a fixed mesh; only the projector matrix that links observations to mesh nodes grows with $N$. At $N = 64{,}000$, SAMBoost completes in 43 seconds, compared with 685 seconds for INLA and 784 seconds for GPBoost with the Vecchia approximation, corresponding to a speedup of roughly a factor of 16-18. GPBoost with exact GP inference is restricted to comparatively small datasets ($N \leq 2{,}000$) due to its cubic time and quadratic memory requirements. INLA is benchmarked in its least expensive practical setting (empirical Bayes integration, no hyperparameter grid search, no CPO), so the reported runtime should be viewed as a lower-bound estimate. The INLA runtime does not increase monotonically with $N$ because, with a fixed mesh, its cost is dominated by internal hyperparameter optimisation whose convergence varies with the data configuration. At larger sample sizes, the data-likelihood cost becomes the dominant term and the scaling stabilises.

\section{Applications to real datasets}\label{sec:application}

\subsection{House price prediction}\label{sec:house}

We assess SAMBoost using the Lucas County house price dataset from Sigrist~\cite{Sigrist2022}, Section~5.2. This dataset (\texttt{spData::house}), originally analysed by LeSage and Pace~\cite{lesage2004models}, contains 25,357 single-family home sales recorded between 1993 and 1998 in Lucas County, Ohio. The outcome variable is the logarithm of the selling price. Covariates cover structural features (age, number of stories, total living area, wall type, counts of bedrooms, bathrooms, half-bathrooms, rooms, and lot size), plot characteristics (frontage, depth, presence of a garage, garage area), and year of sale. Each observation also includes longitude and latitude coordinates.

We adopt the expanding-window temporal evaluation protocol of Sigrist~\cite{Sigrist2022}. For each test year $t \in \{1996, 1997, 1998\}$, the model is trained on all sales from years $< t$ and evaluated on sales in year $t$. Hyperparameters are tuned separately for each fold using an inner split (training on data with year $< t-1$ and validating on year $t-1$), selecting the configuration that minimises validation MSE. This configuration is then retrained on the full training set (years $< t$) and evaluated on the corresponding held-out test year. The resulting training set sizes for the three folds are 11,109, 15,947, and 20,979 observations, respectively. For all three methods, spatial coordinates are used only via the spatial random field and are not supplied to the tree or covariate components. Section~\ref{sec:sim-baselines} details these methods and how they partition variation into spatial and covariate parts. Hyperparameters were selected by grid search under comparable criteria, using 36 combinations for SAMBoost, 27 for GPBoost, and 9 for INLA. The SAMBoost grid was sized to give a tuning budget comparable to that of GPBoost. The SPDE mesh shared by SAMBoost and INLA, with approximately 9{,}000 nodes, was fixed after a preliminary run showed that a coarser mesh missed fine-scale spatial variation, and kept identical across folds. GPBoost employs a Vecchia approximation with 20 neighbours for training and 40 for prediction, together with an unseeded random ordering, which induces slight run-to-run variation (RMSE standard deviation about 0.0003).

\begin{figure}[t]
\centering
\includegraphics[width=\textwidth]{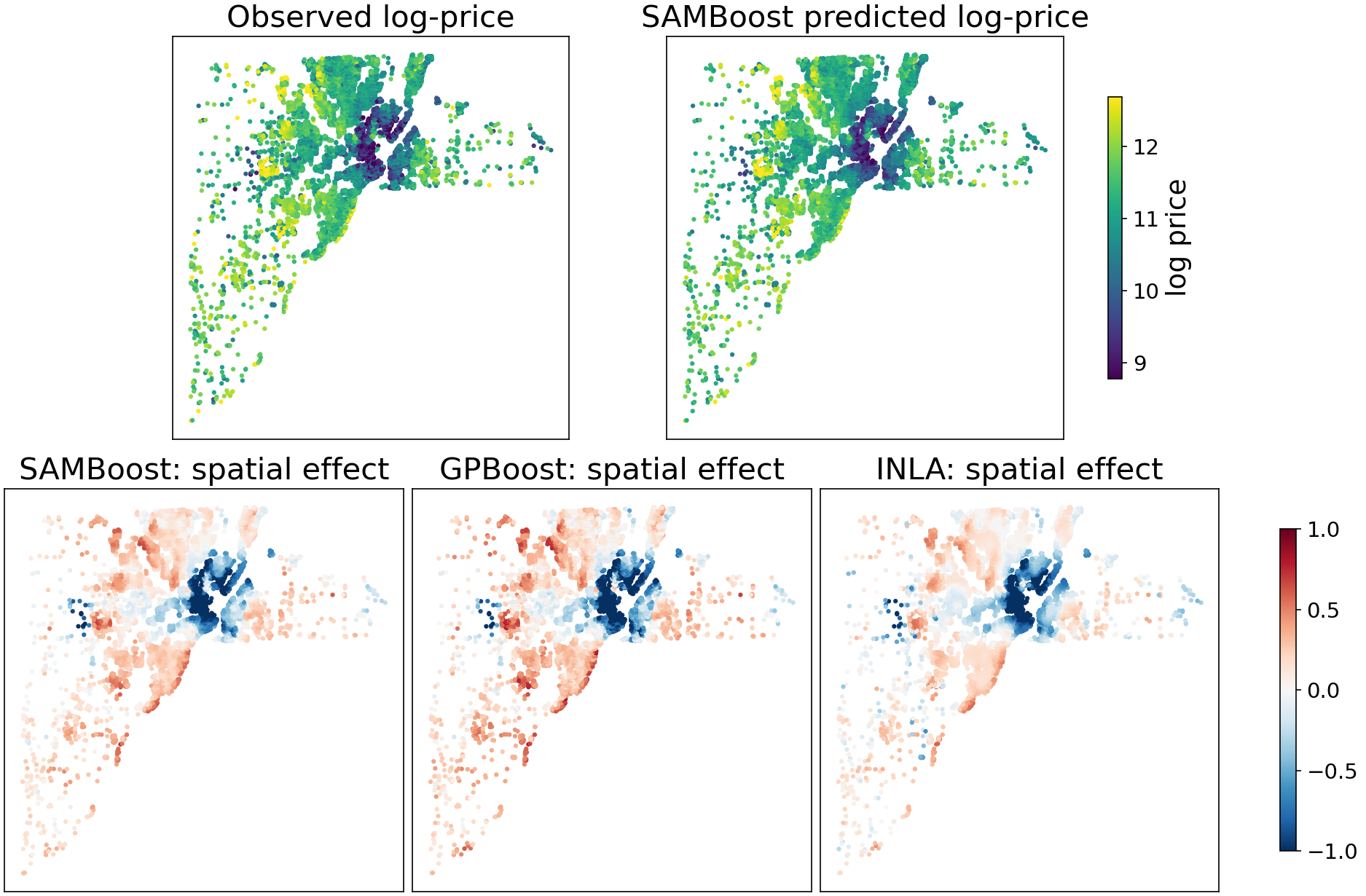}
\caption{Observed and predicted log prices (top row) and estimated spatial effects (bottom row) for the Lucas County data. Only the SAMBoost predictions are shown, as all methods produce visually identical predictions.}\label{fig:housemaps}
\end{figure}

Figure~\ref{fig:housemaps} displays the observed log prices, the log prices predicted by SAMBoost, and the estimated spatial effects for all three approaches. The predicted log prices align closely with the observed ones, and the predictions from the remaining methods appear visually identical to those of SAMBoost, in line with the small RMSE differences reported below. The spatial effects from SAMBoost, GPBoost, and INLA all capture a similar, coherent geographic pattern, with a low-value cluster around the dense urban core and positive effects toward the periphery. This convergence across three different spatial models offers empirical support that the recovered spatial pattern reflects a real geographic signal rather than an artefact tied to any specific method.

Table~\ref{tab:house-results} presents the test set RMSE and mean training time for each method and fold. SAMBoost achieves the best mean RMSE ($0.244$), leading on two of three folds and matching GPBoost on the 1998 fold. A paired bootstrap test~\citep{efron1994introduction} with 10,000 resamples gives a 95\% confidence interval of $[-0.004,\, -0.001]$ for the SAMBoost minus GPBoost RMSE difference, a small advantage for SAMBoost at this mesh resolution. Both approaches substantially outperform INLA ($0.326$). Our GPBoost results also improve on the RMSE of $0.290$ reported by Sigrist~\cite{Sigrist2022} under the same experimental protocol without coordinates in the covariates. This improvement reflects recent software enhancements, which, among other refinements, allow for a smaller Vecchia neighbourhood (20/40 instead of 50/500 in the original setup) without sacrificing accuracy, while markedly cutting computation time. SAMBoost attains performance comparable to that of this updated GPBoost at roughly one third of the fitting time.

\begin{table}[t]
\centering
\caption{Test-set RMSE (log price) and mean fit time on the Lucas County data under the expanding-window protocol. All methods use coordinates only through the spatial field. Best value per column in bold.}\label{tab:house-results}
\smallskip
\small
\begin{tabular}{lccccc}
\toprule
Method & 1996 & 1997 & 1998 & Mean & Fit time (s) \\
\midrule
SAMBoost    & \textbf{0.246} & \textbf{0.239} & \textbf{0.247} & \textbf{0.244} & \textbf{319} \\
GPBoost      & 0.248          & 0.243          & \textbf{0.247}          & 0.246          & 850 \\
INLA         & 0.357          & 0.339          & 0.281          & 0.326          & 1400 \\
\bottomrule
\end{tabular}
\end{table}

To understand the influence of spatial coordinates, we also evaluated SAMBoost and GPBoost with longitude and latitude added to the tree features. This leads to only marginal gains (below $0.0015$ in both cases), suggesting that the SPDE field and GP random effect already capture the spatial structure.

\begin{figure}[t]
\centering
\includegraphics[width=0.75\textwidth]{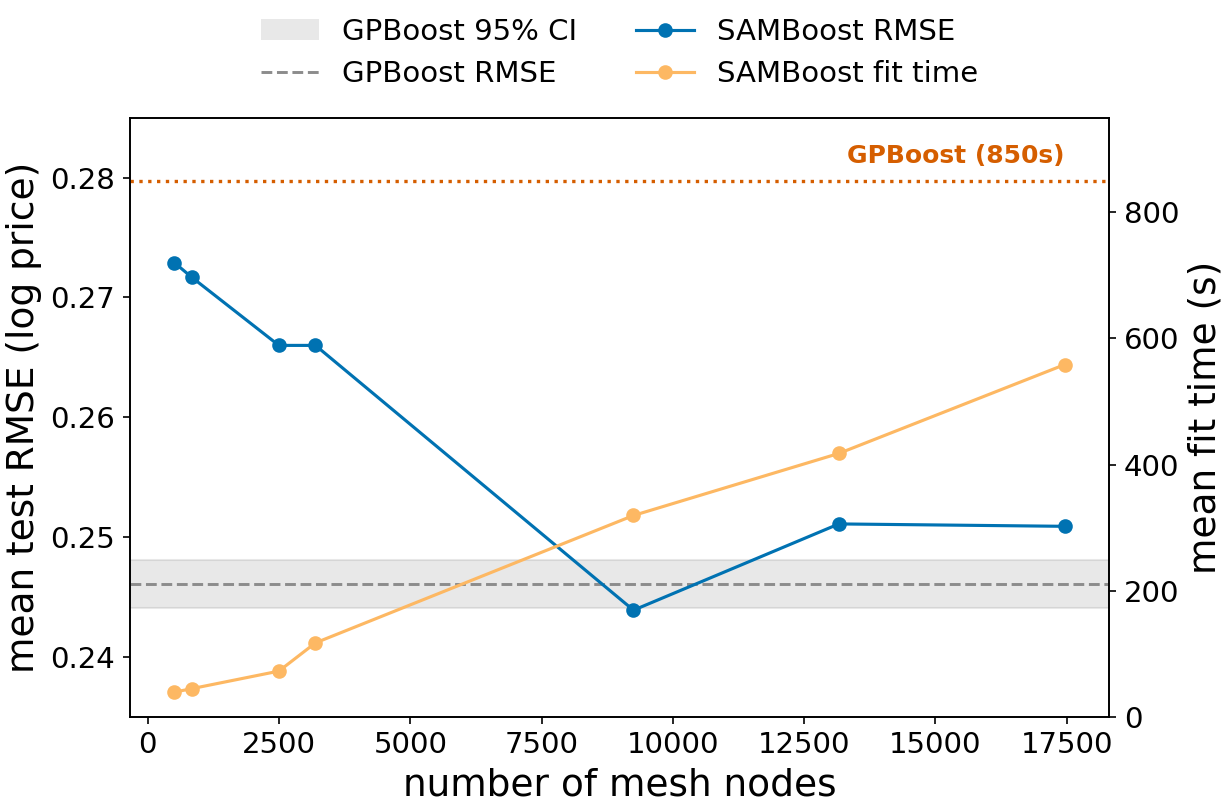}
\caption{Sensitivity of SAMBoost to mesh resolution on the Lucas County data. Mean test-set RMSE (left axis) and mean fit time (right axis) for fixed numbers of SPDE mesh nodes. The grey band shows the GPBoost 95\% bootstrap confidence interval around its mean RMSE (dashed line).}\label{fig:house-mesh}
\end{figure}

SAMBoost defines its spatial field using a finite-element mesh to discretise the SPDE field. An important consideration is how the choice of mesh resolution influences both prediction accuracy and computation time. Figure~\ref{fig:house-mesh} shows the mean test set RMSE as a function of the number of mesh nodes, together with the GPBoost 95\% bootstrap confidence band. With a coarse mesh of 508 nodes, SAMBoost achieves a mean RMSE of $0.273$ in approximately 40 seconds, within 0.03 of the GPBoost RMSE. As resolution increases, the RMSE decreases, reaching a minimum of $0.244$ at approximately 9,000 nodes, where it dips to the lower edge of the GPBoost band. Beyond 9,000 nodes, the RMSE rises again (to $0.251$ at 13,000 and 17,000 nodes). This suggests overfitting, as a finer mesh lets the spatial field fit local variation that does not carry over to the following year. The fit time grows smoothly from about 40 to 560 seconds over the same range. This analysis was carried out after the main experiments and is not used to select the mesh. The resolution used in Table~\ref{tab:house-results} lies near the minimum of the curve, so the SAMBoost results reported there correspond to a favourable but not optimised choice. At lower resolutions, SAMBoost remains competitive at a fraction of the cost, which makes coarse meshes useful for rapid prototyping or exploratory analysis.

\subsection{Flood insurance application}\label{sec:insurance}

We apply SAMBoost to a large French home insurance portfolio covering approximately 968,000 insured properties and 4.9 million policy records across mainland France (excluding Corsica and overseas territories), observed from 2014 to 2022. The objective of the modelling is the occurrence of flood claims, represented as a binary variable ($0/1$) at the policy-record level. The observed frequency of positive cases is about $0.22\%$ (approximately 10,800 claims out of 4.9 million records), so the task is a highly imbalanced rare-event classification problem. The annual positive rate varies by roughly a factor of seven, ranging from about $0.06\%$ in 2017 to $0.41\%$ in 2018, illustrating the sporadic character of flooding. The covariates span underwriting features (number of rooms, declared sums insured, outbuilding size), environmental attributes at the building level (slope, elevation, soil classification, proximity to watercourses), institutional hazard categories, and meteorological metrics derived from rainfall reanalysis products. A thorough account of the covariate construction and an extensive Bayesian spatial investigation of this portfolio are available in earlier studies~\cite{Moriah2026a, Moriah2026}. SAMBoost and INLA share the same SPDE mesh of 12{,}613 nodes, constructed in \cite{Moriah2026}.

The evaluation is based on a temporal hold-out in which models are trained on policy years 2014 to 2019 (about 3 million records) and evaluated on 2020 to 2022. All three methods are compared under this regime. The metrics are exposure-weighted throughout, with log loss as the primary criterion and the Gini coefficient as a complementary ranking measure. Hyperparameters were selected by inner cross-validation on the training set.

\medskip
\textbf{Results.}
Table~\ref{tab:insurance-results} reports the performance on the temporal hold-out set (2020-2022). In terms of log loss, SAMBoost ($0.0297$) and INLA ($0.0305$) perform very similarly, and a paired bootstrap test indicates that their difference is not statistically significant. For the Gini coefficient, SAMBoost ranks first ($0.695$), followed by INLA ($0.660$). In the highest decile of predicted risk, SAMBoost identifies $53.1\%$ of realised claims, versus $50.9\%$ for INLA. Thus, SAMBoost offers a smooth extractable spatial field and achieves essentially the same log loss performance as INLA at a fraction of its runtime, making it the only approach in this comparison that jointly delivers flexible covariate modelling, interpretable spatial structure, and scalability to a portfolio of this size.

GPBoost could not be fitted on the full training set within the four-hour compute budget, even with its fastest Vecchia setup (3 training neighbours). Instead, it was trained on a random subsample of 500,000 rows and evaluated on the same held-out test set as the other models, so its reported metrics reflect the smaller training set. For a comparison on equal data, we also fit both methods at the regional level on a subset of the Occitanie data (197,000 rows, 1,246 positives), in which claims are more frequent than in the region as a whole. SAMBoost attains a higher Gini ($0.774$ versus $0.720$) and a lower log loss ($0.030$ versus $0.034$), although the limited number of positive cases reduces the robustness of this comparison.

\begin{table}[htbp]
\centering
\caption{Flood occurrence results on the temporal hold-out test set (2020-2022). The full portfolio contains approximately 4.9 million policy records, $0.22\%$ positive rate. Log loss and Gini are exposure-weighted. Best value per column in bold.}\label{tab:insurance-results}
\smallskip
\small
\begin{tabular}{lccc}
\toprule
Method & Log loss $\downarrow$ & Gini $\uparrow$ & Fit time \\
\midrule
SAMBoost          & \textbf{0.0297} & \textbf{0.695} & \textbf{119 s} \\
INLA             & 0.0305          & 0.660          & 1996 s \\
GPBoost (500k sub.) & 0.033          & 0.504          & 12,431 s \\
\bottomrule
\end{tabular}
\end{table}

\medskip
\textbf{Scalability.}\label{sec:insurance-scale}
For large sample sizes, the covariate tree stage constitutes the primary contributor to SAMBoost’s computational cost, accounting for approximately 72\% of the runtime per iteration. Switching from the default exact tree splitter to an XGBoost-style histogram-based splitter cuts the per-fit time from 832 seconds to 119 seconds (roughly a sevenfold speedup) with almost no loss in accuracy (Gini $0.695$ vs. $0.697$, log loss identical to the third decimal). The histogram method groups each feature into about 256 candidate cutpoints and uses multi-threading, while the SPDE spatial update is left unchanged. All insurance experiments in this section are based on this histogram-based backend.

Using this backend, a single fit takes about two minutes, which makes repeated fits such as cross-validation over many folds practical, whereas repeating the INLA fit many times is not feasible.

\begin{figure}[htbp]
\centering
\includegraphics[width=\textwidth]{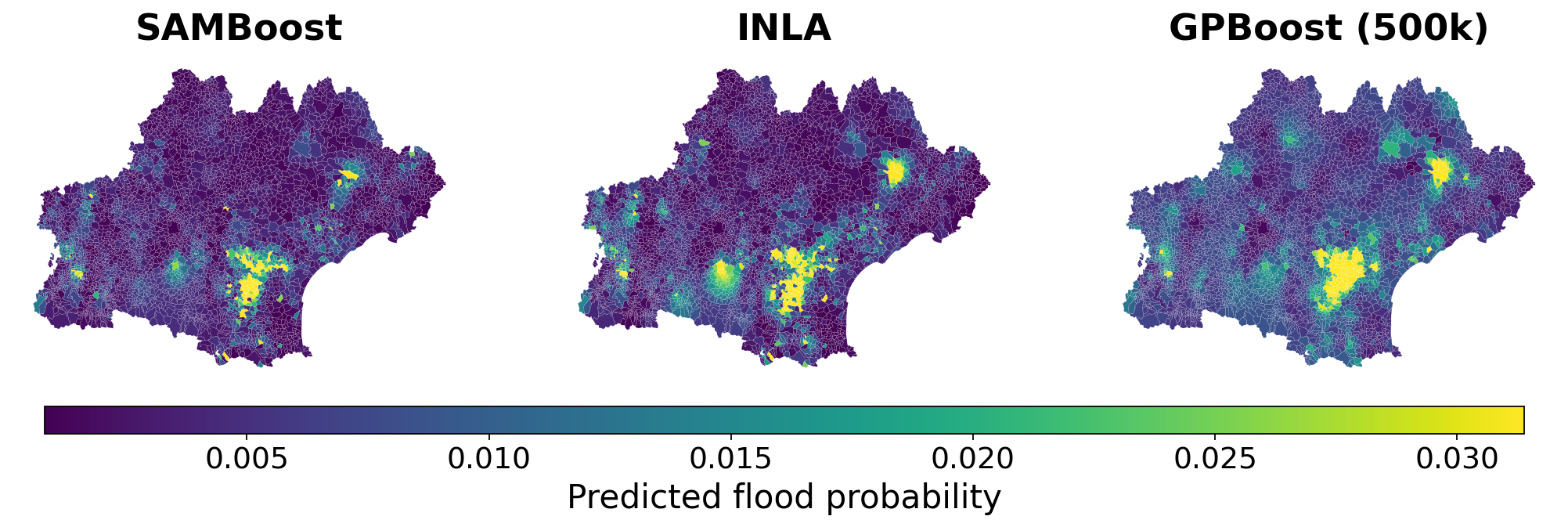}

\vspace{0.3em}

\includegraphics[width=\textwidth]{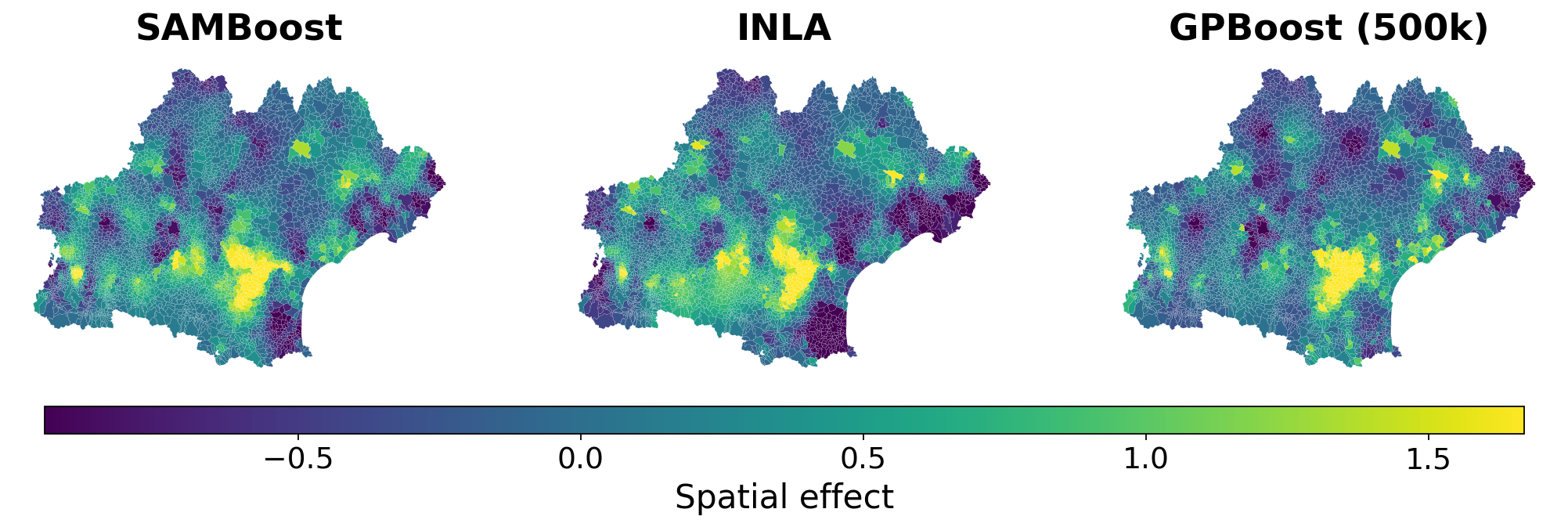}
\caption{Predicted flood probabilities (top) and estimated spatial effects (bottom) over the Occitanie region. GPBoost is trained on a random subsample of 500,000 records.}\label{fig:insurance-maps}
\end{figure}

\medskip
\textbf{Spatial effect maps.}
Figure~\ref{fig:insurance-maps} shows the predicted flood probabilities and the estimated spatial effects for the three methods, displayed over the Occitanie region (the only region shown for confidentiality reasons; all numerical results refer to the national portfolio). The predicted probability surfaces from SAMBoost and INLA identify the same broad risk pattern, with higher probabilities clustered along major river corridors and in regions classified with high institutional hazard levels. While GPBoost’s predicted probabilities are influenced by the subsampling artefact described above, its spatial pattern remains qualitatively similar.

The spatial components of SAMBoost, INLA and GPBoost reconstruct a coherent geographic surface of comparable amplitude, with positive effects over the established flood-prone areas (notably the Aude and Hérault valleys) and negative effects where flood risk is low.

\medskip
\textbf{Spatial feature augmentation.}
The methods compared above all estimate an explicit spatial field. The simplest practical alternative, described in Section~\ref{sec:intro-review}, instead appends the coordinates to the covariates of a standard learner, which produces no spatial field that can be reported separately. We kept it out of the main comparison for this reason, but its predictive performance remains a natural reference, both as a baseline in the spatial boosting literature~\citep{Sigrist2022} and in insurance pricing practice~\citep{HenckaertsCoteetal2021}. We therefore fitted XGBoost~\citep{ChenGuestrin2016}, a widely used gradient boosted tree implementation that remains a strong reference on tabular data~\citep{GrinsztajnOyallonVaroquaux2022,ShwartzZivArmon2022}, on the covariates augmented with longitude and latitude, using the same tuning and evaluation protocols as the other methods. It is less accurate than SAMBoost in three of our four settings, with an AUC of $0.648$ versus $0.689$ in Simulation~1 and an RMSE of $0.635$ versus $0.596$ in Simulation~2 and $0.259$ versus $0.244$ on the Lucas County data, where Sigrist~\cite{Sigrist2022} also found boosting with coordinates close to GPBoost in accuracy. On the flood portfolio, it achieves the best scores, with a log loss of $0.0291$ and a Gini of $0.717$ against $0.0297$ and $0.695$ for SAMBoost, both differences being significant. This gap comes from space-covariate interactions that the additive structure of SAMBoost cannot represent, since adding the raw coordinates to its tree features raises its Gini to $0.716$, at the cost of the decomposition. XGBoost is also the fastest method in all our experiments, with $36$~s versus $119$~s for SAMBoost on the flood portfolio. The price of an explicit spatial field is therefore a longer fitting time and, on the flood portfolio only, a small loss in ranking accuracy.

\section{Discussion}\label{sec:discussion}

Spatially structured data often create two demands, flexible covariate modelling and an explicit spatial field that can be reported on its own. Spatial statistical models such as INLA meet the second demand but require the covariate effects to be specified in advance, while tree-based methods meet the first but either have no explicit spatial field or treat it mainly as a correction for the covariate effects rather than as an object of interest. SAMBoost addresses both demands within a single boosting loop, for applications in which the spatial field is itself a quantity of interest. It was built around three design goals, namely an explicit spatial field, flexible covariate effects, and scalability to large datasets. We review each goal in light of the results, together with its limitations.

The first goal is an explicit spatial field that is separated from the covariates. When the covariate model of INLA is adequate, the fields recovered by SAMBoost are close to those of INLA. Compared with GPBoost, predictive differences are small, but the spatial field is recovered markedly better ($\mathrm{rmse}_{\mathrm{spa}}$ $0.160$ versus $0.207$ in Simulation~2), which supports giving the field the same iterative refinement as the covariate model. In insurance, for instance, this field serves as a geographic risk surface in its own right, and the mesh resolution and $\lambda_{\mathrm{spa}}$ control how local it is, and hence the balance between risk differentiation and mutualisation. The main limitation is that SAMBoost returns a point estimate of the field, whereas INLA provides its full posterior distribution. A Gaussian approximation built from the Newton system at the final boosting round could provide approximate uncertainty, but this remains to be explored.

The second goal is flexible covariate modelling. When covariate effects include interactions, the additive model of INLA cannot represent them and the unmodelled residual degrades both components (its $r_{\mathrm{cov}}$ falls from $0.924$ to $0.625$ in the sensitivity analysis), whereas SAMBoost remains stable. When the true effect is close to additive, INLA performs as well and additionally provides posterior inference, so SAMBoost is most useful when the covariate structure is complex or unknown in advance, as is typical of observational datasets. The exact separation also has a price. Effects that depend jointly on location and covariates cannot be represented by either component, which explains the Gini gap with XGBoost on the flood portfolio. Adding the coordinates to the tree features closes this gap but gives up the decomposition, which is the purpose of SAMBoost. This cost remains small ($0.022$ in Gini) and appears only on the flood portfolio, the most complex of our applications.

The third goal is scalability. SAMBoost fits the full flood training set in about two minutes, against more than thirty minutes for INLA with a simplified setup, while GPBoost could not be fitted on the full training set. The main remaining cost lies in tuning. The SPDE range parameter $\kappa$ is selected by cross-validation, whereas INLA and GPBoost estimate their spatial hyperparameters internally. The number of boosting rounds was also chosen by grid search without early stopping, so the reported fit times are not optimised. Multi-resolution scheduling, with coarse meshes in early rounds and finer ones later, could further reduce the cost of each fit.

Overall, this work shows that a latent spatial field can be estimated as an explicit component within gradient boosting, with a recovered field consistent with Bayesian spatial models, flexible covariate effects, and a computational cost compatible with large datasets such as national insurance portfolios. Beyond the directions discussed above, the separation between a tree ensemble and a mesh-based latent field offers a general template for combining machine learning with structured latent effects.

\begin{Backmatter}

\paragraph{Acknowledgments}
We are grateful to ADDACTIS France for providing access to the insurance data used in this study. We also thank Fabio Sigrist for helpful correspondence on the GPBoost implementation.

\paragraph{Competing Interests}
None.

\paragraph{Data Availability Statement}
The Lucas County house price data are openly available in the R package spData. All data used to build the explanatory variables of the flood application, including the French natural disaster database (GASPAR), are openly available. The insurance portfolio cannot be shared for confidentiality reasons. The SAMBoost implementation and the code to reproduce the experiments are available at \url{https://github.com/MoriahMulah/SAMBoost}.

\bibliography{bibartun.bib}

\end{Backmatter}
\end{document}